# IMPLEMENTING BOOLEAN FUNCTIONS WITH SWITCHING LATTICE NETWORKS

by

Rajesh Kumar Datta
B.S., Chittagong University of Engineering and Technology, Bangladesh 2014

A Thesis
Submitted in Partial Fulfillment of the Requirements for the
Master of Science Degree

Department of Electrical, Computer and Biomedical Engineering
in the Graduate School
Southern Illinois University Carbondale
August 2021

**THESIS APPROVAL**

IMPLEMENTING BOOLEAN FUNCTIONS WITH SWITCHING LATTICE NETWORKS

by

Rajesh Kumar Datta

A Thesis Submitted in Partial

Fulfillment of the Requirements

for the Degree of

Master of Science

in the field of Electrical and Computer Engineering

Approved by:

Dr. Dimitrios Kagaris, Chair

Dr. Haibo Wang

Dr. Chao Lu

**AN ABSTRACT OF THE THESIS OF**

Rajesh Kumar Datta, for the Master of Science degree in Electrical and Computer Engineering, presented on June 24, 2021 at Southern Illinois University Carbondale.

TITLE: IMPLEMENTING BOOLEAN FUNCTIONS WITH SWITCHING LATTICE NETWORKS .

MAJOR PROFESSOR: Dr. Dimitrios Kagaris

Four terminal switching network is an alternative structure to realize the logic functions in electronic circuit modeling. This network can be used to implement a Boolean function with less number of switches than the two terminal based CMOS switch. Each switch of the network is driven by a Boolean literal. Any switch is connected to its four neighbors if a literal takes the value 1 , else it is disconnected. In our work, we aimed to develop a technique by which we can find out if any Boolean function can be implemented with a given four-terminal network. It is done using the path of any given lattice network. First, we developed a synthesis tool by which we can create a library of Boolean functions with a given four-terminal switching network and random Boolean literals. This tool can be used to check the output of any lattice network which can also function as a lattice network solver. In the next step, we used the library functions to develop and test our MAPPING tool where the functions were given as input and from the output, we can get the implemented function in four terminal lattice network. Finally, we have proposed a systematic procedure to implement any Boolean function with a efficient way by any given one type of lattice network.

.

## ACKNOWLEDGMENTS

I would like to thank Dr. Dimitrios Kagaris for all his help and guidance through my education. He has been a great influence on my interests and without his help, it would have not been possible to complete this thesis. He has developed a major part of this thesis. I am grateful to him for all his suggestions and considerations in my graduation journey. I would also like to thank Dr. Haibo Wang and Dr. Chao Lu for being members of my thesis committee and for their valuable suggestions. Finally, I would like to thank the College of Engineering for the amazing learning experiences I have had during my time here at Southern Illinois University, Carbondale.

.

## LIST OF TABLES

TABLE PAGE

## LIST OF FIGURES

FIGURE PAGE

# CHAPTER 1

# INTRODUCTION

In the semiconductor manufacturing industry shrinking of the dimension of CMOS transistors has played a key role in the last few decades. But this shrinking trend which was predicted by Gordon Moore in 1965 (Moore Law) is going to end in the near future[1]. To overcome this problem, four-terminal switching network concept can be a very useful technology to replace the conventional CMOS based circuit. This network can be used to implement the same Boolean function with less number of switches than the two terminal-based CMOS switch [2]. Circuit modeling and layout technology with four terminal switching has been studied in recent time [9][15]. Four terminal switch based network modeling offers great area advantage over the regular two terminal based switch modeling. Two terminal switches have two states - ON(closed) and OFF(open). A Boolean function can be implemented with these two-terminal switching networks in series/parallel configuration. Each switch has a Boolean literal as input. If a literal value is "1" then the corresponding switch is ON and if "0" the switch is OFF. Computation is done by taking the product of the literals in each path.

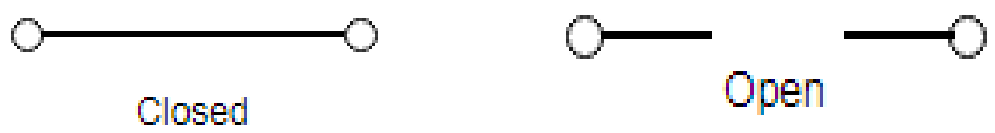

Fig1: States of Two Terminal Switch

In a switch all the four terminals are either mutually On or mutually Off. These networks are considered to be arranged in rectangular lattice. We can also call them 'Switching lattice' . Every rectangular switch is controlled by a Boolean literal. If a literal is "1" then the corresponding switch is ON and if "0" the switch is OFF. The states are same as two terminal switch but now the connections can be made at the four side of a rectangular switch.

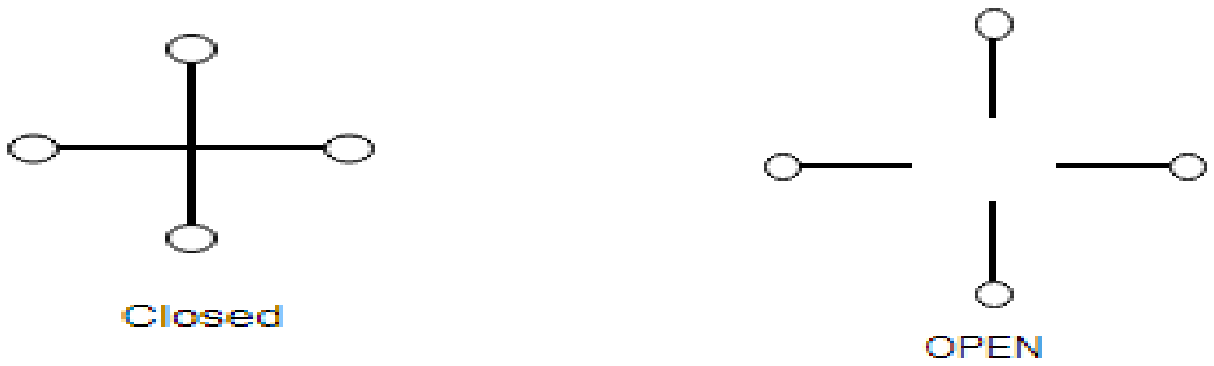

Fig 2: States of Four Terminal Switch

Implementation of a Boolean Function in a four terminal network can be done by the following the below rules:

- A Boolean literal will control every switch. If a literal is equal to "1" the switch can be connected with its four neighbor. Otherwise it is not connected.
- If any input assignment produces a connected path between top and the bottom edges and evaluates to 1, that will be a valid path in the network. It can be considered as a Product term of the Boolean Function. If the path evaluates to 0, it does not exist. For instance, We can take a Network of 3 by 2 switching lattice network as shown in Fig 3.

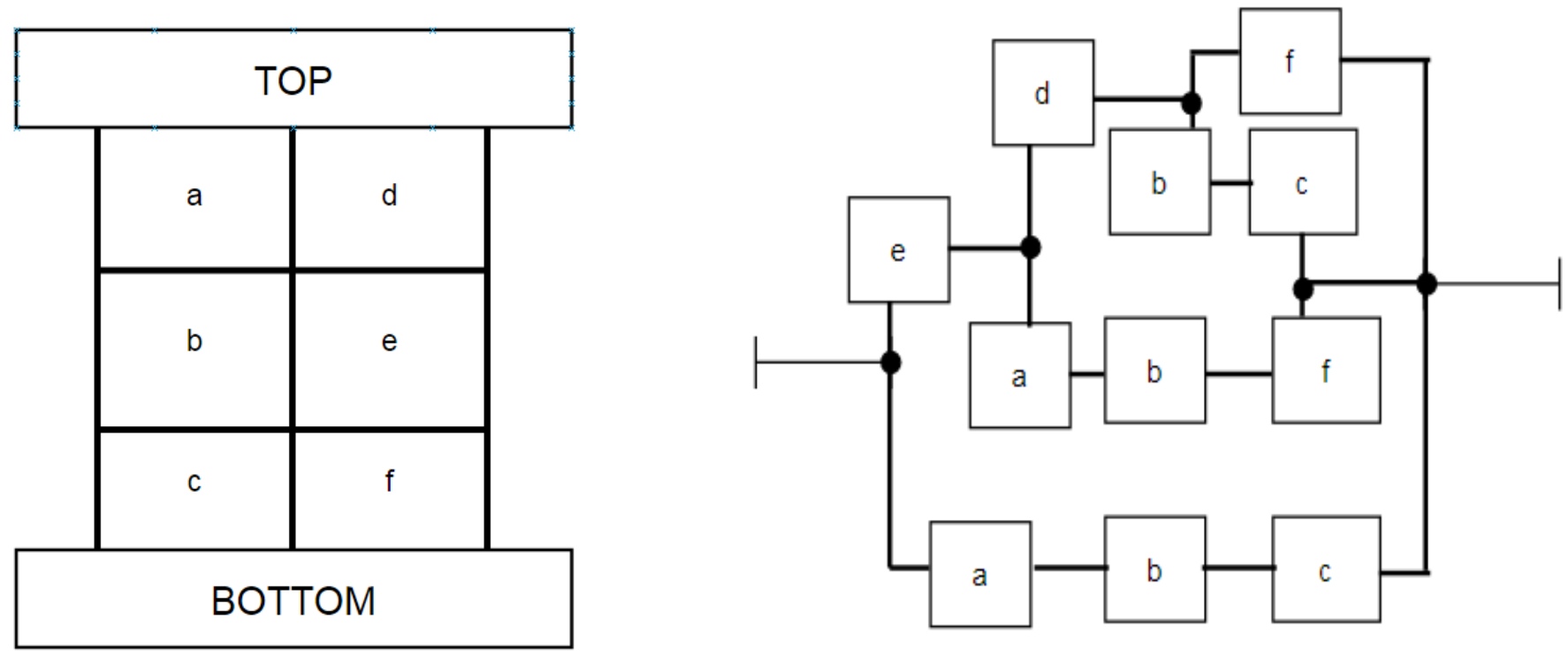

Fig 3: Implementation of function X = a b c+ a b e f+ d e b c + d e f with (a) Four terminal (b) Two terminal switch

The implementation of the function X = a b c+ a be f + d e b c + def is shown in Fig 3(a) with four terminal lattice. This is a much more compact implementation of the Boolean Function than with regular two terminal switch shown in Fig 3(b). Only 6 four terminal switches are required for this implementation . Each product term of the function is one path from the TOP to BOTTOM of the lattice.

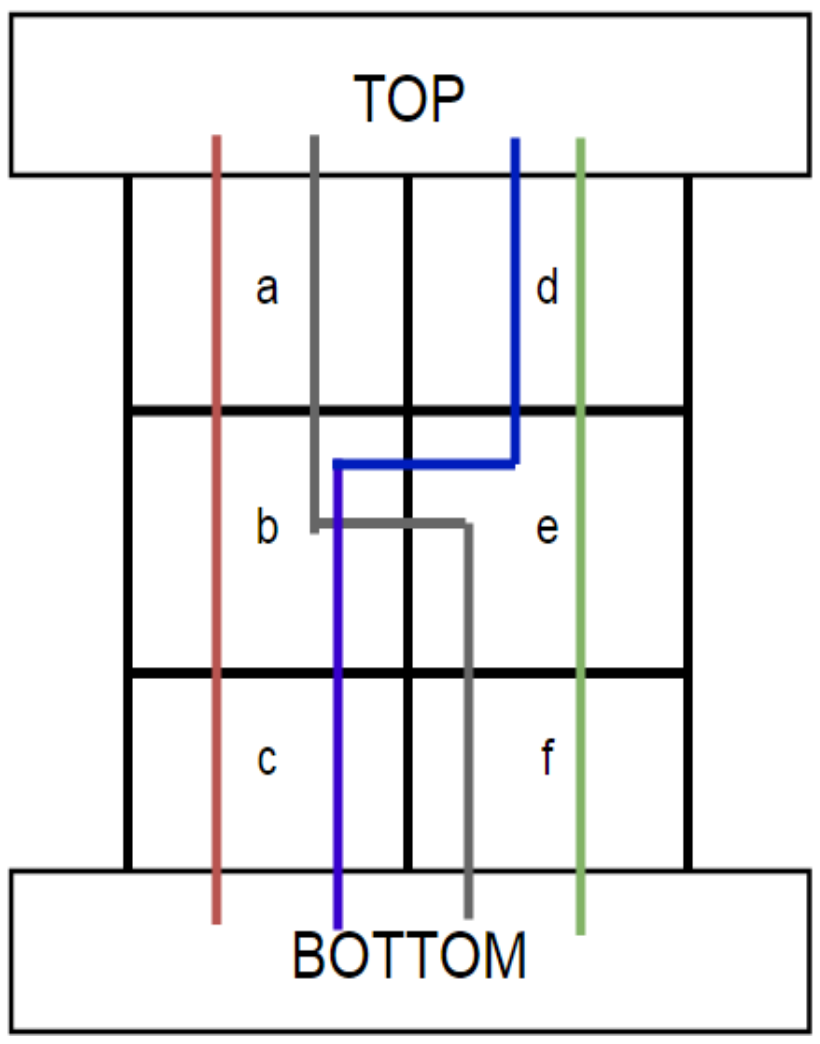

Fig 4: Lattice implementing X = a b c+ a b e f+ d e b c+ d e f

Fig 4 shows the detailed path connection of the function X = a b c+ a b e f+ d e b c+ d e f. Here each line connected from TOP to BOTTOM is a Product term. Details about how this lattice formation is done and how this works is mentioned in [2]. Our aim was to develop a process by which we can find out if any Boolean Function can be implemented with a given four terminal network . The result will be the final literal assignment of the lattice network.

## CHAPTER 2

## METHODOLOGY

### 2.1 Related Works of Four terminal networks

The concept of two dimensional arrays of a four terminal switch is not new . In a seminal paper (1972) Akers introduced about this model [4]. With new technologies this model has renewed attention in recent years [5],[6]. Logic function can be achieved by crossbar type switches[7],[8]. Using all these concepts four terminal network model was proposed in [2]. Fig 5 shows the schematic of the four terminal switch concept. Every rectangular lattice is a four-terminal switch. Switches are controlled by high (logic 1) and low (logic 0) voltage. Logic 1 means switch is ON and 0 means switch is OFF. We will consider the top-to-bottom connectivity across the network. If top and bottom plates are connected the output signal will be logic '1' and if not the output will be logic '0'. From this model , output can be taken with a resistor that is connected to the bottom. At the Top a high voltage need to be applied. Technologies have been studied to fit this model in [2][15]. Realization of the function mapping in lattice network using SAT solver was studied in [13] [17].

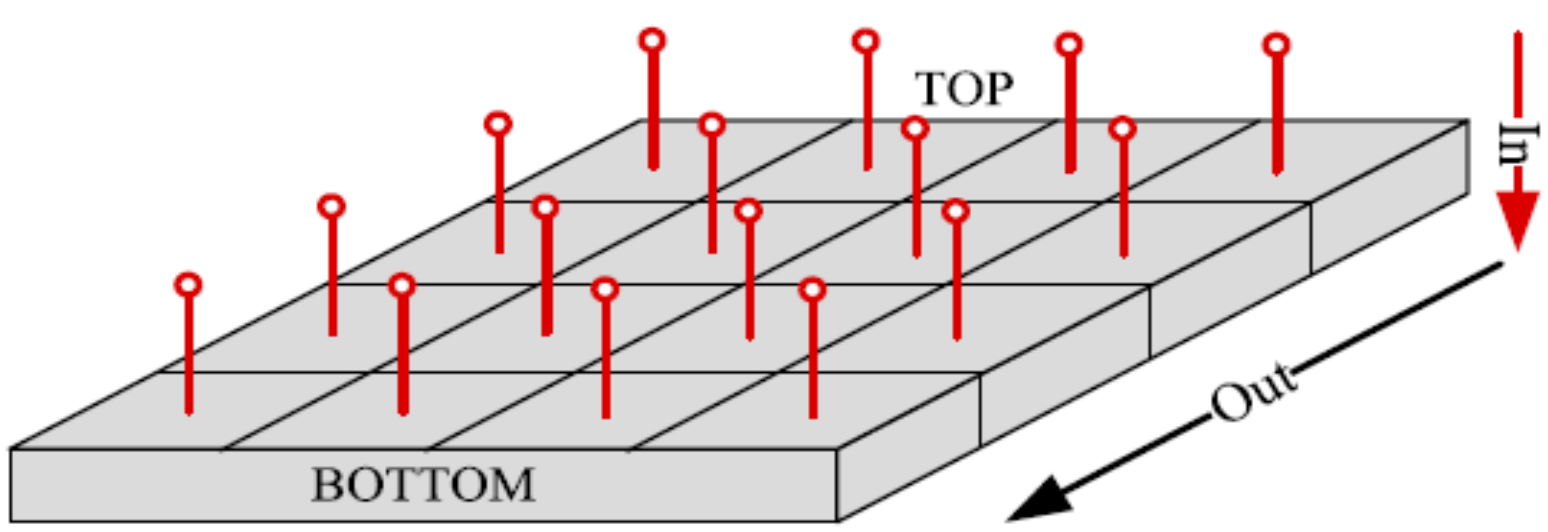

Fig 5: 3D realization of our circuit model with the inputs and the output

## 2.2 Formation of a Lattice

The formation of the four-terminal network and its features are discussed in this section. We can take an example to build a simple lattice at first. For instance, we can take a lattice of 3 by 3. Here, first '3' means the number of rows in the lattice and the second '3' is the total column number. So in this network, we have a total of '9' four-terminal switch. The upper part is considered as 'TOP' and the lower part is considered as 'BOTTOM'. Fig 6 shows this configuration.

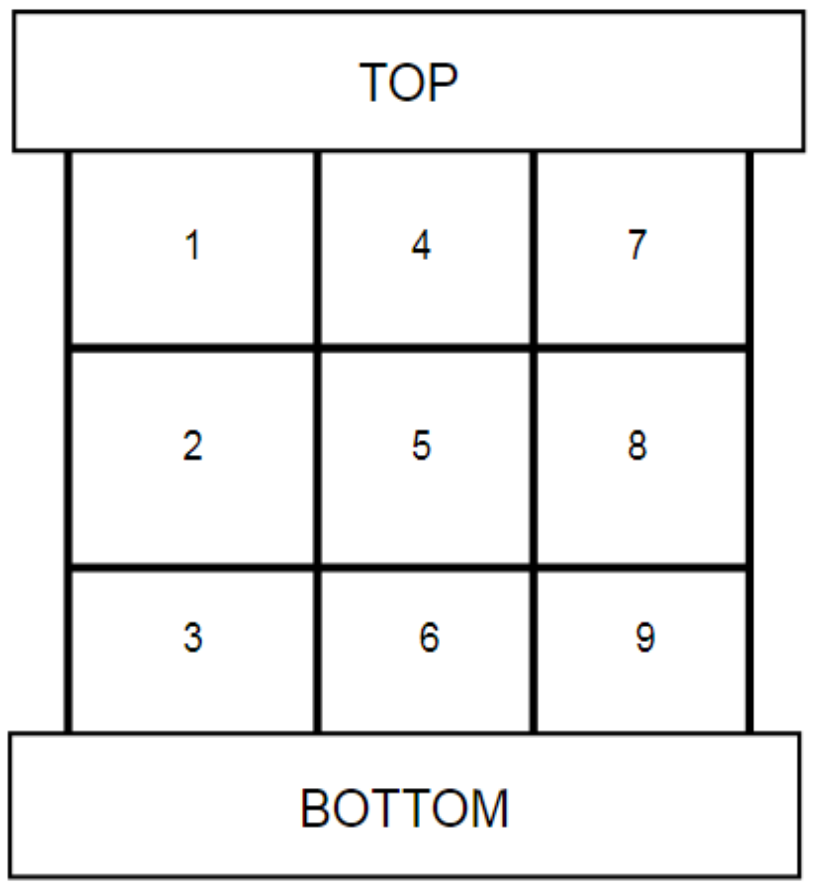

Fig 6: 3 by 3 lattice configuration

Every path that is connected from top-to-bottom is considered to be a Product term of a Boolean function. For now, we are considering every lattice has a different literal. Though all lattices are connected, we have to consider only the valid paths. Here valid paths are those which does not have any superset. We can consider an Example to check which one is a valid path and which is not.

from Fig 6, we can take following two paths:

TOP -> 1-> 2-> 3 -> BOTTOM

TOP -> 1-> 4-> 5-> 2-> 3 -> BOTTOM

Both of these paths connect the top and bottom plate. Now we replace them with Boolean literal (a, b, c, d.... etc). From '1 to 5' , we take each number as literal from 'a to e' .Thus these two path becomes:

a->b->c

a->d->e->b->c

Here second path is the superset of the first path as all the elements of the first path is present at the second path. Now, simply we can write these two paths as 'a b c' and 'a d e b c'. These are two Sum of Product of the four terminal network. But in a function they will be considered as ' a b c + a d e b c '. Clearly we can write this as 'a b c (1+ d e)' . Doing Boolean algebra we can write this term as 'a b c' ( In Boolean algebra : A + 1 = 1 ) .

From the above example, we can see that all the paths that connect Top to Bottom cannot be considered as the valid sum of product. We have to remove any superset paths that will be produced. For the 3 by 3 lattice of Fig 6 , following are the only valid paths:

7 8 9

7 8 5 6

7 8 5 2 3

4 5 6

4 5 2 3

4 5 8 9

1 2 3

1 2 56

1 2 5 8 9

We have got total 9 paths .

For our design, it is necessary to find out only the valid paths for any kind of lattice formation we have. If we do not repeat any 'literal' as input of switches we can get the highest number of possible paths a lattice can build with any four terminal structure. These paths increase rapidly if our lattice size increases.

### 2.2.1 Literal Selection rules

For our design, we have considered some rules for assigning literals in the four terminal lattice network. Literals will be represented as numeric number at the input of every switch. We will consider all the 26 letters. For representing 'a' we will use '0', for 'b' we will use '1'. And thus we will use numbers from '0' to '25' for representing all the letters. So if we have a sum of product 'a d e' then it will be presented as '0 3 4' at our output.

We also need to consider the complement of a literal. For complement of a literal, we will take '1000 - literal'. For Example, if we want to take the complement of the literal 'a' which is the complement of numeric value '0', the complement of 'a' will be '1000-0' or '1000'. So, a' (complement of 'a') is represented by 1000. Similarly, b' (complement of 'b') will be represented by 999, c' (complement of 'c') will be 998, and so on. For Example, if we have an SOP " a b' c ", it will be represented with " 0 999 2". We have to consider the constant ' 1 ' value. It will be represented with ' 101 '. Constant ' 0 ' will be represented by ' 100 '.

### 2.3 Boolean Function Representation

With the proper literal selection in our four terminal model, we will create the library of Boolean function that can be implemented with any given four terminal network. We will represent the output of a Boolean function as follows:

Let's consider a function : a b c d' + c d b' + d b c' + b c d e f

Now with our literal selection process we will choose

a= '0'

b= '1'

c= '2'

d= '3'

e= '4'

We also have to take the complement of these literals . So we will select '1000-x' for the complement of the literals. Where 'x' is the non complement value of the literals.

a' = 1000- '0' = 1000

b' = 1000- '1' = 999

c' = 1000- '2' = 998

d'= 1000- '3' = 997

For representing every product term we will first use the total number of literals the product has. We will represent every product term in a single line. Thus our Boolean function can be represented as follows:

Function : a b c d' + c d b' + d b c' + b c d e f

Function representation:

4 0 1 2 997

3 2 3 999

3 3 1 998

5 1 2 3 4 5

Here we can see, first product term has 4 literals. The first element of the product term is 4 which means this product has 4 literal. Rest part of the line is our first Product term ( a b c d'). Thus Basically above representation means the Boolean function which is

4 a b c d'

3 c d b'

3 d b c'

5 b c d e f

## 2.4 Design Approach For Basic Lattice Model

### 2.4.1 Lattice dimension

For the design of the tool first, we need to take the lattice dimension. A lattice consists of rows and columns. We can have a four terminal network of m by n dimension. Here m is the row of the lattice and n is the column. Suppose we want to test a lattice of 3 by 3 network. So it has 3 number of rows and 3 number of columns. In this structure, we have to use 9 four terminal switch. Fig 7 shows the structure of the network with 9 connected switch .

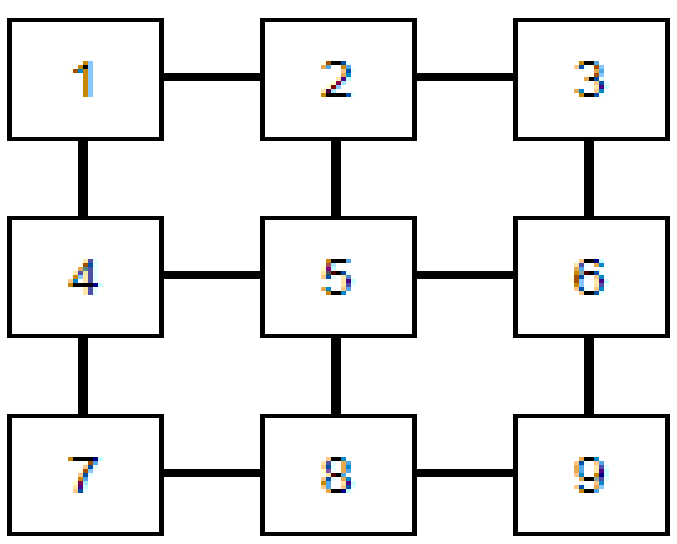

Fig 7: 3 by 3 four terminal network with 9 connected four terminal switch

In our design, we have taken input from the user about the row and column. If the user gives the input our tool will generate the basic lattice as per the given dimension. This basic network needs to have a TOP and Bottom connection. In our design we will place a 'Source ' at the TOP and 'Destination' at the BOTTOM. These will work same as the Top and Bottom plate of a switching network.

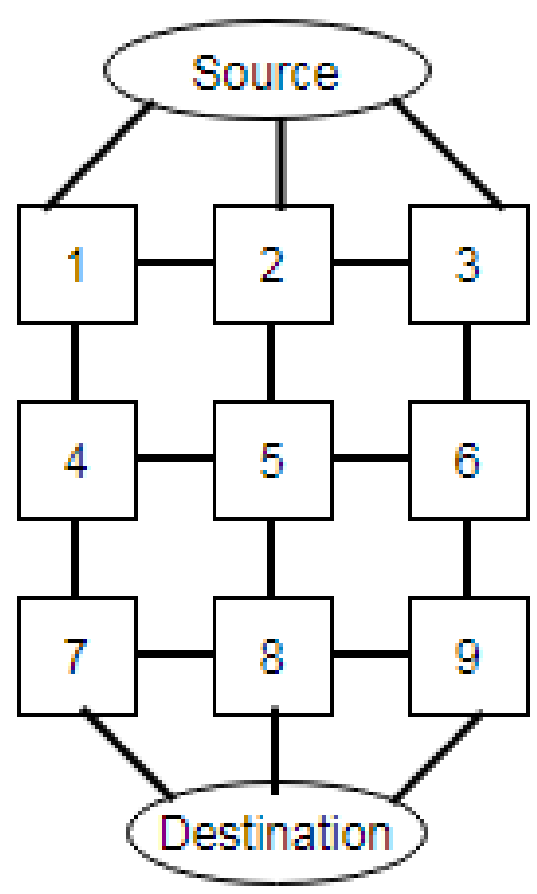

Fig 8: 3 by 3 four terminal network with Source and Destination

Fig 8 shows the network with Source and Destination. Source and Destination is connected with the switches of upper row and lower rows of the network. A valid path or a valid product term should start from the source (Top) and ends at destination (Bottom). So we need to check the paths that connects the source and destination. This lattice network can be considered as graph data structure. We can consider each switch as a node of the graph. For finding the paths we need a search algorithm. In our design we have used the Depth-first search (DFS) algorithm for this purpose. With this algorithm we can find all the valid paths that are connected from source to destination . But all these paths are not valid. We need to remove all the non valid paths (Superset paths) from our design.

### 2.4.2 Modification in Lattice Configuration

In our design we do not need 'supersets' of other paths to be generated. We had to find some characteristics of the switching lattice by which we can easily stop the paths to be generated during execution of DFS. Firstly, we can easily reduce some useless paths by changing the connections between the switches in the switching lattice. If we take the connection of the Fig 8, which is a basic four terminal network, let us consider 2 paths which connects the source and destination.

source -> 1-> 4 ->7 -> destination

source -> 1-> 2-> 5-> 4 ->7 -> destination

Here clearly second path is the superset of the first path.

Also we can check another two paths.

source -> 1-> 4 ->7 -> destination

source -> 1-> 4 -> 7 -> 8 -> destination

Here also second path is the superset of the first one.

These kinds of paths which are generated because of the 'right' side connections of the four terminal switches at the first row and last row of the lattice network can be prevented by changing the connection like Fig 9.

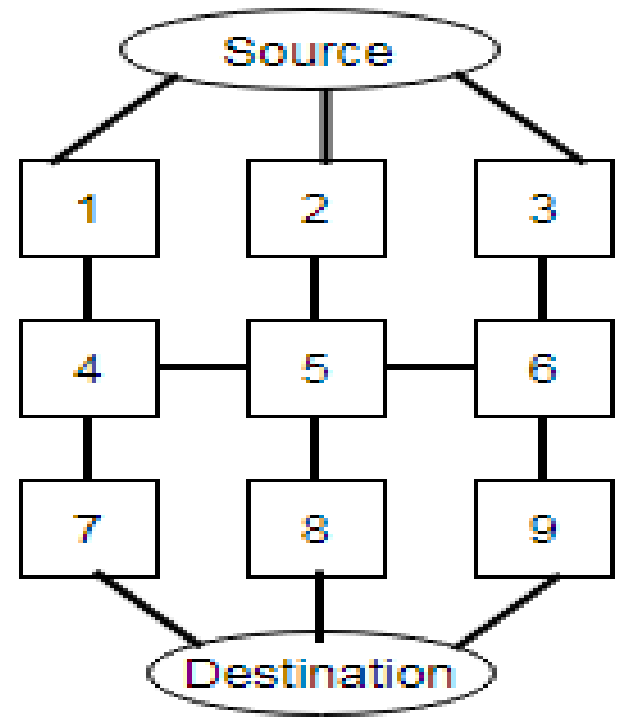

Fig 9: Reconfigured 3 by 3 four terminal network

This new configuration will prevent generating supersets we mentioned in our Example in this section previously. With this change, we can prevent some useless paths in our path generation process. If the lattice goes bigger these useless paths increase rapidly. So this change will make our design tool faster than with the previous configuration. In Fig 10 we represent another lattice form (4 by 4) of this configuration.

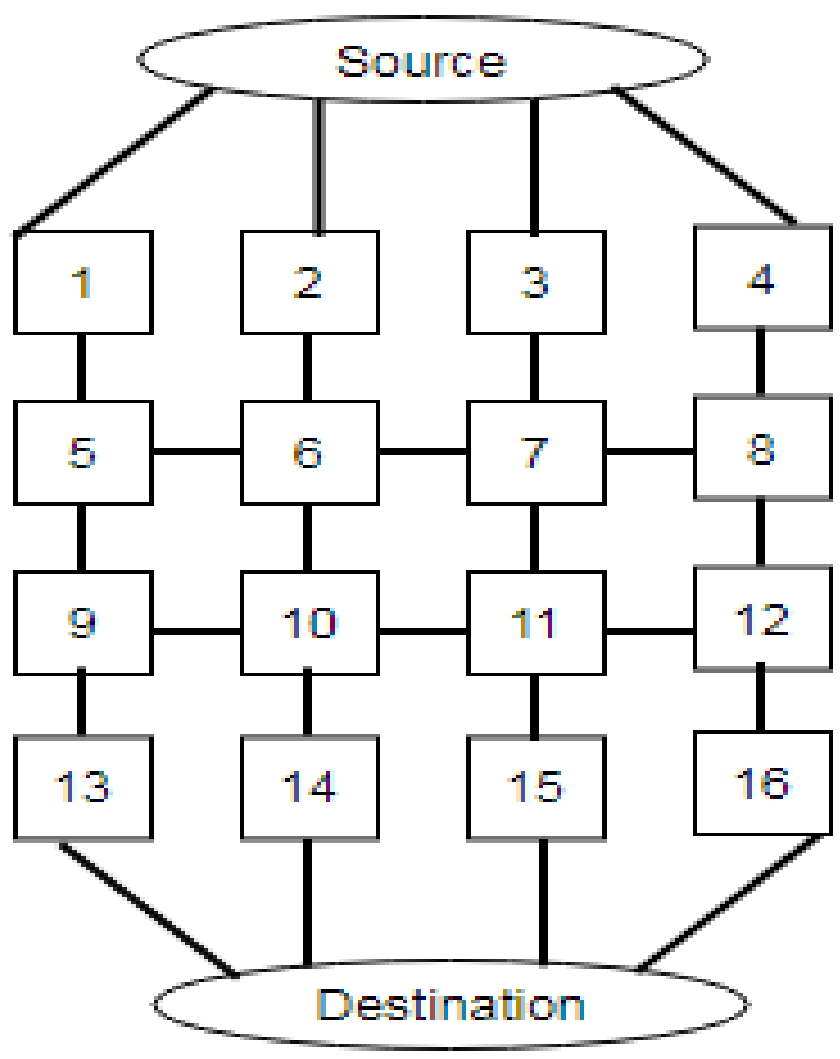

Fig 10: 4 by 4 switching network

### 2.4.3 Path Formation

For generating paths, we first check the adjacent node of a single four-terminal switch. We will call the adjacent nodes 'Children' of the main node. We know every four-terminal switch can be connected with 4 adjacent switches. So one four-terminal switch can have maximum 4 children..

Let's consider the lattice model of Fig 10 and check the children of the nodes. If we take switch '1', we can see it is connected with two nodes. One is the 'source' and the other is '5'. This is the same for all other switches in the 1st row of the network. Similarly, in the last row, every switch has 2 children. Switch '15' has children as 'destination' and 11. If we take switch '10', we can see it has 4 children ( 6, 9 , 11, 14). Thus we can find out all the children for every node. It is required for our path generation process. We need to form a 'Children' array which will consist of the children nodes of every switch.

Now , we can explain with Example how one path is generated. We will use the network of Fig 10 to explain. Our aim is to begin from 'Start' and following are the required steps:

- First check the children of the source. From the Figure we can see, source has for children "1 , 2 , 3 , 4".
- Now we have to pick any one child of the source. Let us pick '1'. Next we will check the children of '1'. It has one child '5'. '5' has two child '6' and '9'. We can choose any one to proceed. We choose '9'.
- Following same procedure, we get '9' has children '10' and '13'. From this we choose '10'. '10' has children '11' and '14'. From these, we choose '14' for our next step. '14' has one child that's 'Destination'. Here we have reached to our destination.
- So now we have a valid path . Our path is :

  " Start -> 1 -> 5-> 9-> 10 -> 14 -> Destination"

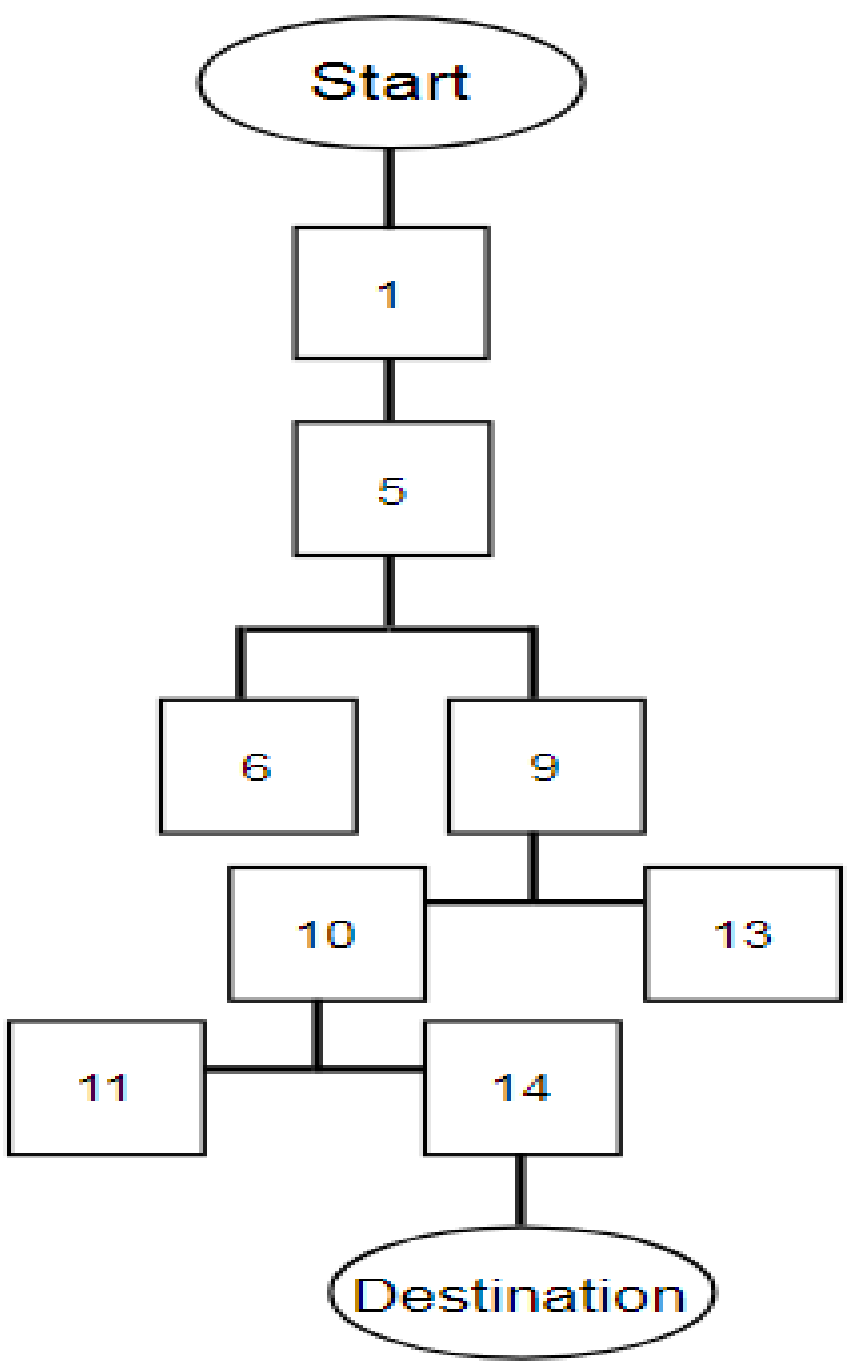

Fig 11: Formation of a path from 'Source' to 'Destination'

This process will be repeated recursively till all the nodes and their children are examined. If one child is already used in previous paths, it will be marked. It will not be checked during the next search of a new path. Thus we will get all the paths. With Depth First Search (DFS) algorithm this process can be implemented..

### 2.4.4 Removal of Supersets

After generating a path we need to check if it is a superset of any other path. A way to do this can be generating all the paths first and then by checking if any path is a superset of another path. Then we can remove the paths which are a superset. But this way is very much time-consuming. The code will take a long time to generate all the available paths first and then removing supersets. The main reason for this delay in this process is the rapid growth of the paths with the increasing lattice size.

For making our design process faster we will check a path during its generation. There is a special feature of this four-terminal network that can be used to easily remove invalid paths. Let's explain the logic with an Example with a 4 by 3 four network model.

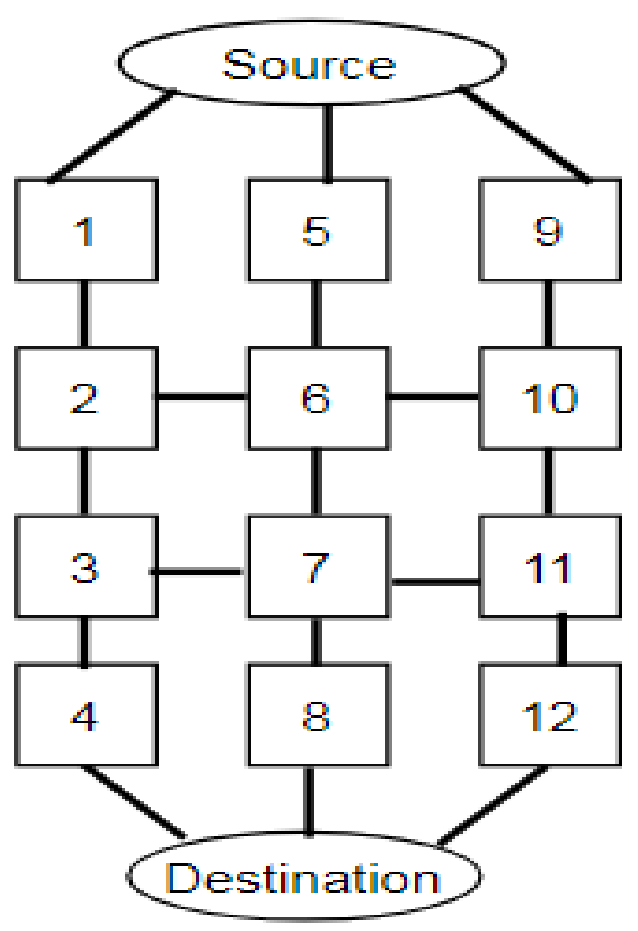

Fig 12: 4 by 3 Four terminal network

Let's take two paths one of which is superset of the other path.

1->2->6->7->3->4

1->2->3->4

Here, 1st path is the superset of the second path. We want to design our process so that, the first path which is the superset of the second path will not be generated at all. We will cancel this path during the path generation process.

Following is the logic we will apply in our design :

**"** If in a currently developed path a new node 'Y' (lattice switch) is added we will check if that node is a child of any PREVIOUS node in the currently developing path excluding the Node 'X' that "brought" Node ' Y' in the path**. "**

In the Example given in this section " 1->2->6->7->3->4 " has 5 nodes and it is a superset path. If we apply our logic in this path, we can see 1, 2, 6, or 7 is not a child of any other previous node except the immediate node. If we check '6 '; it is the only child of '2 ' in this path. It is not a child of '1'. Similarly '7' is the only child of '6'. Not child of '1' or '2'. But if we see '3' from the lattice diagram we can see, '3' is a child of both the previous node '7' and also the node '2'. So from this, we can say this path will be a superset of any other path. So we will remove it before the path is generated. If we take ' 1->2->3->4 ' , we can see '2' is a child of only '1', '3' is a child of only '2' and '4' is a child of only '3'. So this path is genuine and we should keep that. So applying this logic we can easily remove the supersets before they generate from our design. This way is super fast and takes very little time than the way that was mentioned at the beginning of the section. We have implemented our design with this logic.

If we combine and implement all the logic till now we will get the valid basic paths of the four-terminal switching network. Notice that, till now we are considering all our lattice switches have different inputs. With our design, we can find out how many paths we can have with basic r by c structure .

In Table 1 we can see how the product terms (paths) increase with the growth of the lattice size .

Table 1: Number of product terms with different lattice structure.

| **r/c** | **2** | **3** | **4** | **5** | **6** | **7** |
|---|---|---|---|---|---|---|
| **2** | 2 | 3 | 4 | 5 | 6 | 7 |
| **3** | 4 | 9 | 16 | 25 | 36 | 49 |
| **4** | 6 | 17 | 36 | 67 | 118 | 203 |
| **5** | 10 | 37 | 94 | 205 | 436 | 957 |
| **6** | 16 | 77 | 236 | 621 | 1668 | 4883 |
| **7** | 26 | 163 | 602 | 1905 | 6562 | 26317 |

## 2.5 Placement of the Boolean Function literals

Up to section 2.4, we implemented basic lattice dimension design rules by which we can determine the maximum possible valid path (product ) in a lattice. Now in these four-terminal switches, we need to assign Boolean literal inputs. And with those given Boolean literals we will get the Boolean Function that a lattice structure can implement. Following the rules of Boolean literal selection explained in 2.2.1 we have to select the literals. The paths we got from the basic lattice formation are the maximum possible paths with the lattice. Now in a basic path, each element will be replaced with Boolean literals. We need to consider the following points. Literals can be repeated in a valid path. Constant

- '1' means this term will not appear in the product term but it will not make any change to the other elements of the product. It will be represented as '101' in the input of the Boolean literal. Constant '0' means it will cancel the whole path. It will be represented as '100' in the input of the Boolean literal.
- If any path has a literal and also the complement of that literal then the path will be canceled.

If in the final output of any lattice there is only 1 product term, the count of the literal of that product term is 1 but there is no literal, it means the answer is 'Constant 1' only.

For the generation of the library of the function which is one of our design goals, we will select a range of the literals from which we can choose any literal randomly for every switch. The maximum range will have 25 literals starting from 0. Also, we need to consider the complement of the literals which we have chosen. And along with all the literals and their complement, we also include '101' and '100' in the range.

How to make a lattice range we can easily explain with an example. With the design of basic path formation that has been described up to 2.4, we will have a design tool that can give us the output of all the valid paths from 'source' to 'destination'. Now at this point, we will take the Boolean literals as input when we have a valid path. Let's consider we have a 3 by 3 model of the lattice dimension. Up to 2.4, we have a design that will generate a valid path that connects Top and Bottom. Now in the design, when we get a valid path we will replace that path with the literal inputs which will be taken randomly as input of those switches during lattice formation.

Suppose, we have taken literal range from 0 to 4 for our 3 by 3 lattice model. So this range basically will represent 'a to e'.

[0, 1, 2, 3, 4] = [a,b,c,d,e]

Now we will take the complements in this range.

[1000, 999, 998,997, 996 ] = [a', b' , c', d', e']

We also have to include 101 [constant 1], 100[constant 0] in the range.

So the final range from which we take literal for the lattice is as below :

[0, 1, 2, 3, 4, 1000, 999, 998,997, 996,101, 100]

From this range randomly literals will be assigned for the four terminal switch inputs. For instance , let's say in random selection we have picked following 9 literals from the range to be the input of the 9 switches in 3 by 3 lattice network randomly.

Let's consider the selected literals to place in the 3 by 3 four terminal switches are: [0, 1 , 2, 101 , 1, 2, 1 , 998 , 101]

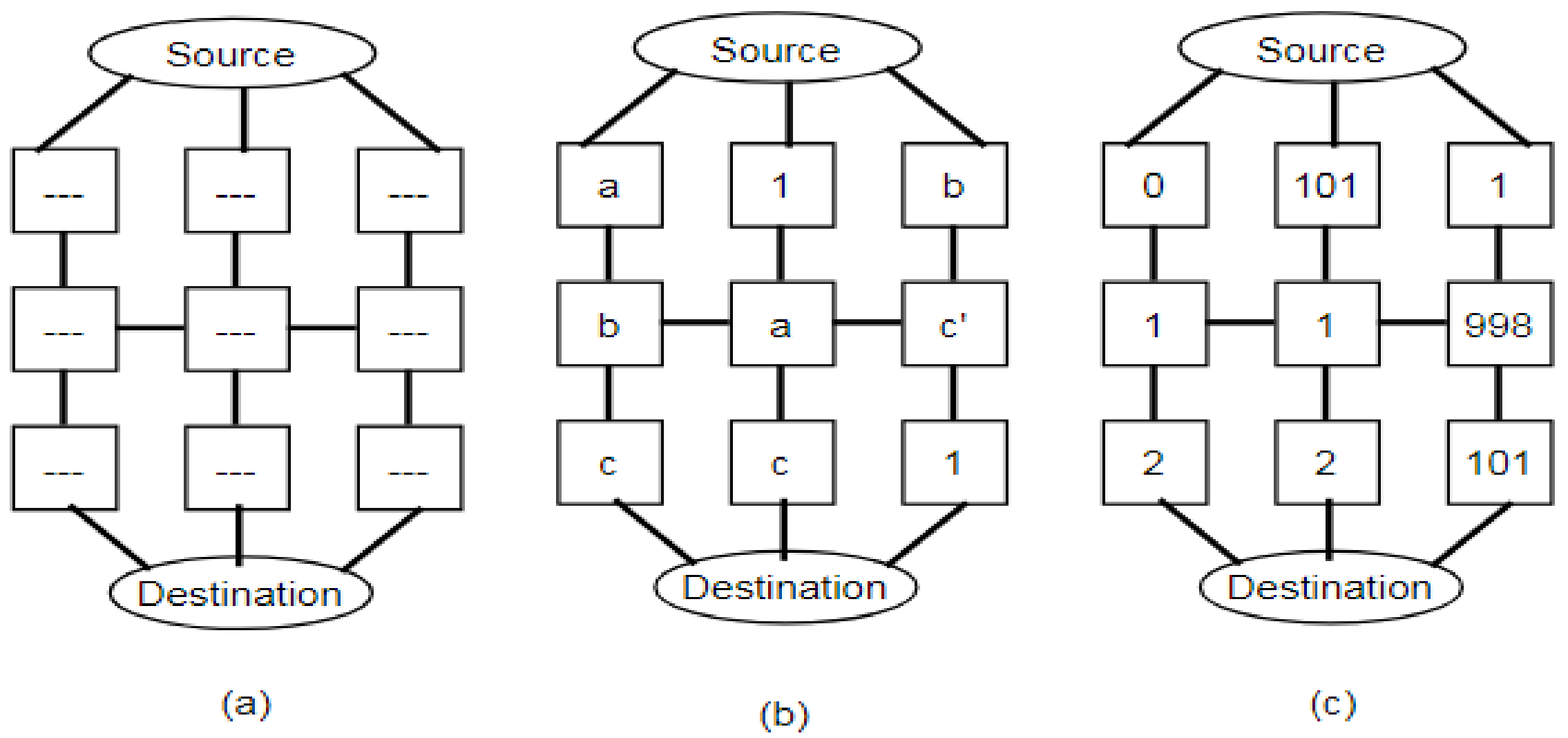

Fig 13 (a) A basic 3 by 3 lattice formation (b) lattice input with literals (c) lattice input with literals for the design tool

Fig 13 (c) is the final four terminal network with the literal values. Now we need to consider the paths that are valid for this network with the literals. Here we have to consider the supersets again. Because previously we removed the supersets from the basic lattice model generated paths(considering all the switch has different inputs). We removed those supersets and got the results that are the valid paths for basic lattice. But now this lattice with literal values can also have supersets, as we can see there are literals that can be repeated more than once. Also if in a path there is element '101', then we have to consider it as constant '1' and then we have to search for any superset.

For example, from Fig 13 (c) we can take 2 paths from source to destination.

source -> 0 -> 1 -> 2-> destination

source -> 101 -> 1 -> 2-> destination

Here the second path has element '101' which is basically 'Constant 1'. This means the path is actually ' source -> 1 -> 2-> destination' which is the subset of the first one. So we can say that 'source -> 0 -> 1 -> 2-> destination' is a superset among these two paths. So this superset should not be included in the final product term. Again if there are any same literal more than once is a path, we will not consider more than once. For example, "source -> 0 -> 1-> 1 -> 2-> destination" This path is basically "source -> 0 -> 1 -> 2-> destination" . So if we have the below paths,

source -> 0 -> 1-> 1 -> 2-> destination

source -> 101 -> 1 -> 2-> destination

first one should be removed and the second one should remain as

' source -> 1 -> 2-> destination'

So for this removal of the supersets, we need to use some separate design logic that will find out the supersets and eliminate them. We need to implement logic that will check every path with all other paths. If there is any path that has all the elements of any other path and the second path has fewer literals in total than the first one, then the first path will be a superset path. From our output of the Boolean function that path should be eliminated.

For the lattice showed in Fig 13(c) using all the logics we described we will get the following outputs.

2 1 2

2 1 998

If we check the outputs as a Boolean function we can represent them as a sum of products.

So the function will be:

$f = b\,c + b\,c'$

In this case if we see further elaboration we will get ,

f = b( c + c') = b [using Boolean algebra law]

so the final result we should get

1 1

## 2.6 Library of the Functions

To make the library of functions we have to execute the above mentioned process for several times. In the design we can choose how many times we want the trial for generating different functions. If we choose a trial number , every time random literal will be chosen from the literal range that we set earlier. For Example , if we set the trial number for 20 times, the whole process will execute for 20 different times and 20 Boolean functions that can be implemented with the given lattice model will be generated. Every time Boolean literal will be selected randomly and will produce different Boolean function. Thus we can build the library of the functions for the given lattice model with our design tool. The output of a lattice will be formed as follows. One sample output is given as follows.

```
3 3                  // row and column of the lattice
---------------------------------------------------------------------------
1000   998   1000
101     0     1               This is the lattice   with Boolean literals
1       999   1
---------------------------------------------------------------------------

2          /// total number of product terms that can be implemented  with the lattice
---------------------------------------------------------------------------
2  998  0                          ////generated product terms ///////

2 1000  1
---------------------------------------------------------------------------
```

So the final output will represent " c' a + a' b "

## 2.7 Lattice Network Solver:

The function generation process was done with random Boolean literals. The tool we developed can be used as a Lattice Network Solver too by doing some modification. Instead of giving random literals if we give the literal of any circuit implementation to the circuit, it will give the function implementation of the circuit. This can be used to check if we have implemented any function with the lattice structure correctly. In the checking of a big lattice structure, it can be very useful. Some studies have been done on the fault model of switching lattice networks [16]. This tool can be used to diagnose any fault in the lattice implementation.

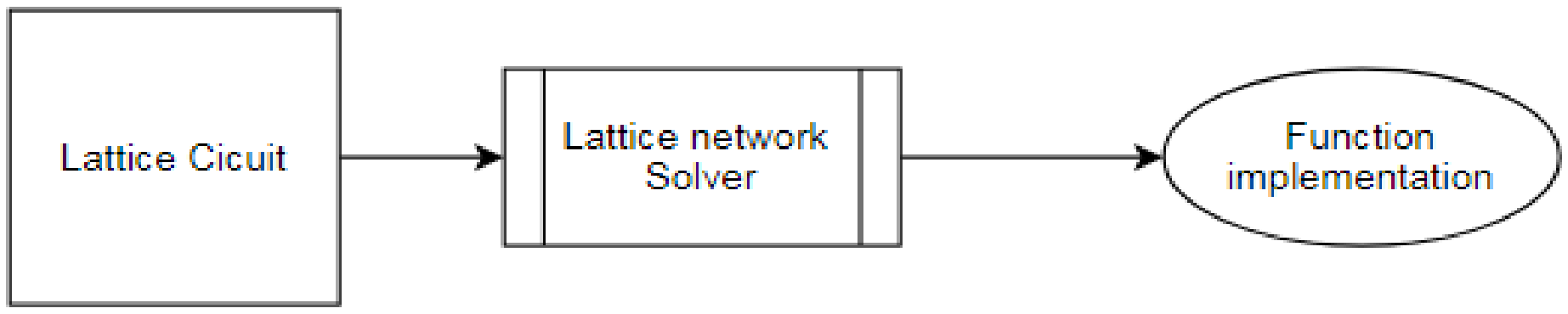

Fig 14: Lattice network solver overview

For Example : if we have a complex circuit implementation with 6x6 lattice it will have a lot of Product terms. Finding the output of the lattice will be so difficult manually.

| 11 | 15 | 991 | 3 | 12 | 13 |
|---|---|---|---|---|---|
| 5 | 986 | 992 | 992 | 15 | 985 |
| 995 | 1 | 2 | 100 | 8 | 0 |
| 998 | 8 | 987 | 990 | 996 | 990 |
| 10 | 997 | 2 | 994 | 14 | 993 |
| 986 | 990 | 3 | 986 | 0 | 988 |

Fig 15 : Example of a 6x6 Lattice

It will difficult to solve the function output for this lattice. So if we run it through our tool we can get the output of the function. This function will not be minimized. It will give the exact implementation result of the circuit. Following is the result for this circuit implementation with lattice network solver:

```
Lattice Graph
11 15 991 3 12 13
5 986 992 992 15 985
995 1 2 100 8 0
998 8 987 990 996 990
10 997 2 994 14 993
986 990 3 986 0 988

  number of nodes 38
```

Fig 16: (a) A 6x6 Lattice network circuit

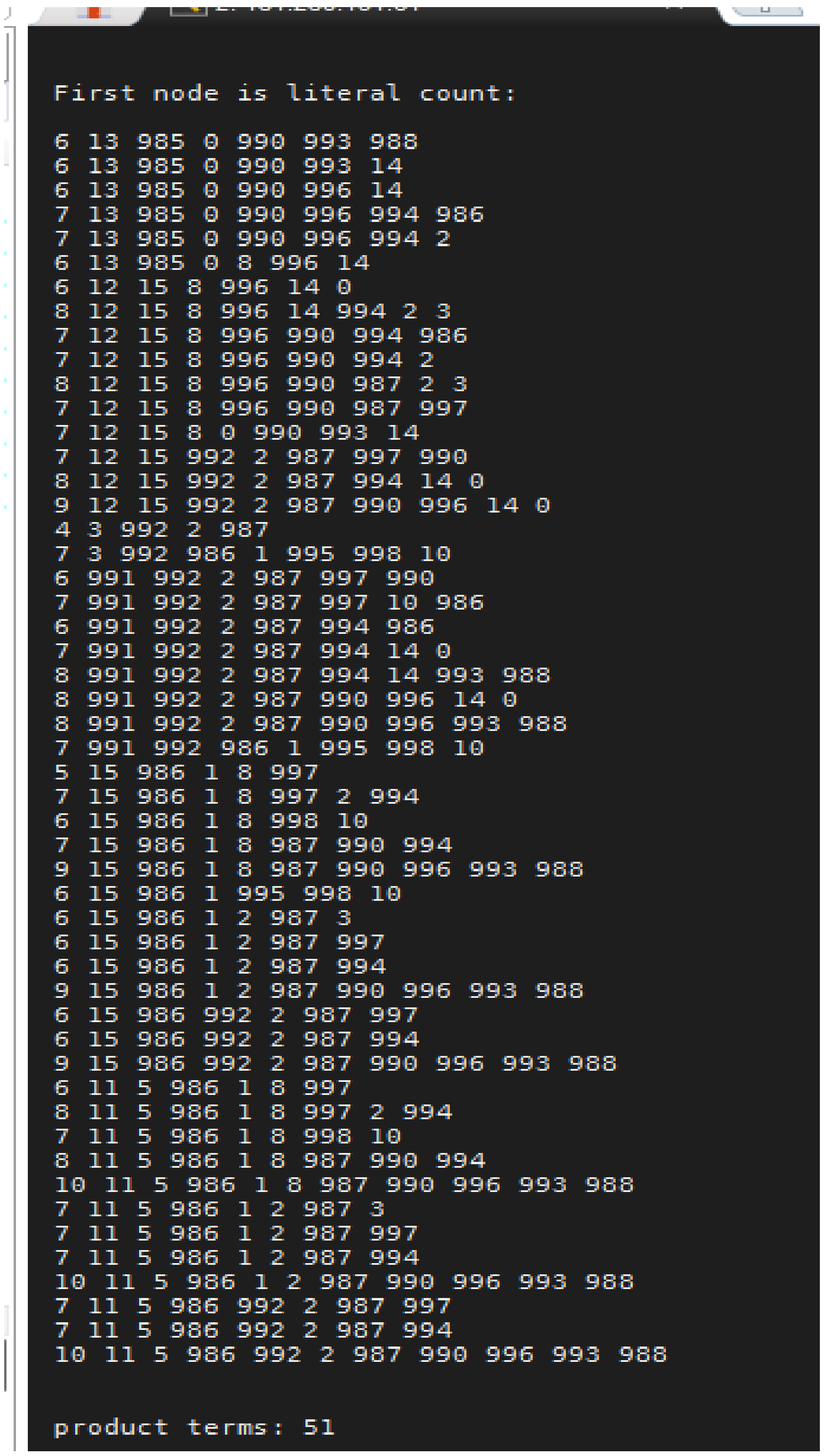

Fig 16 (b): Output of the Lattice with 6x6 with the lattice solver tool

The output has 51 product terms means this circuit will have 51 product terms in the form of Sum of product. The number of product terms were matched with the original output.

# CHAPTER 3

## MAPPING OF BOOLEAN FUNCTIONS IN LATTICE

In the previous chapter, we discussed how we can generate a library for Boolean functions from any given dimension of switching lattice. These functions are the output with random variables and these are the Boolean Functions generated from the Switching lattice. Now we can use them to test the Mapping tool which was the ultimate goal for this work.

### 3.1 MAPPING TOOL

To achieve the final goal of our work, we need a MAPPER which can map a Boolean function in a switching lattice network if it is possible to implement the function with the given Lattice network. If we give product terms of a BOOLEAN FUNCTION as input of the MAPPING tool, the tool will try to MAP the product terms using the path of the given lattice network. If it can map all the terms the output of the lattice will be the implementation of the given function. The library functions that we generated from our previous tool, now can be used as input of the MAPPING tool to test the correctness of the tool. Those functions are generated from switching lattice networks and we can use them to test if our MAPPER can map those functions with the lattice network.

The MAPPER will use the basic structure of a switching lattice network to map any given function. For Example, if we take a 3 by 3 switching lattice network , it can have maximum 9 paths (If all the four terminal switch in the lattice network has different Boolean literals).

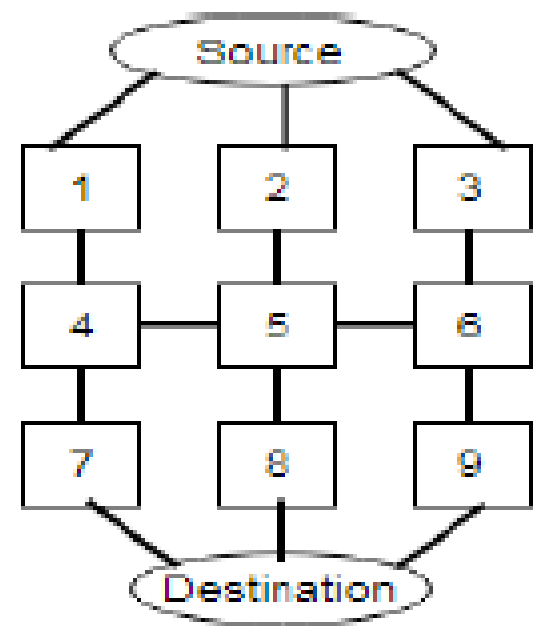

Fig 17: 3 by 3 switching lattice network

Paths from a 3 by 3 switching lattice network:

It has 9 paths and 9 literals in total.

9 9
5 0 3 4 5 8
4 0 3 4 7
3 0 3 6
4 1 4 5 8
4 1 4 3 6
3 1 4 7
5 2 5 4 3 6
4 2 5 4 7
3 2 5 8

With these basic paths we need to map a function given as inputs to this lattice network. For example: if any path has 3 product terms, MAPPER will try to search the combination of the lattice formation which will give the output of that lattice as our input function.

### 3.2 MAPPING tool overview

The mapping tool was developed by **Dr. Dimitri Kagaris**. This is a compact tool to find out the lattice formation of any function. We don't need to use any minimization tool or SAT solver for this process. We will also show a flowchart after giving the short description. Also, the output of the sample results will be analyzed. .

Inputs of the mapping tool will be given by the user . There will be two inputs . One is the paths of a given lattice network and other will be the product terms of the given input function. We can generate path of any lattice terminal by using the rules of the lattice network in chapter 2. Also we need to convert any function to the input form as we mentioned for the functions in previous chapter. Using this two inputs the mapping tool will try to house the literals of all the product terms in the path of the lattice network . We will mention the mapping tool as MAPPER in our writing. In this section, we will first mention the overview of the process and then present the flow chart.

Following are the basic steps overview that we need to follow:

STEP 1: Take the lattice network paths and the product terms of the target function as input.

STEP 2: Create the group of the lattice network paths according to their size. Paths will be used according to the Cardinality.

STEP 3: The Product terms will be tested with different examination orders. One order will be send as a input first, and if necessary then the next examination order will be generated. It will continue to check different order of inputs until we got a solution or all the examination orders are checked.

STEP 4: Start the matching process with the examination order of STEP 3.

STEP 5**:** Check if we have available (not used before) paths (shortest first). If path is available then go to the next step. Else check if we have already got the solution .If there is no solution, check for the next examination order of the input by going back to STEP 4 (if all the orders are not examined already), otherwise there is no solution for the function.

STEP 6: Select an available Path from the 'Path Group ' (Shortest first). Try to match the product term with the selected path. If all the literals of the product term can be assigned in the path without any conflict with any previous assignment, then that product term can be housed in that path. Go for the next Product term (If any term is remaining). If there is conflict or cannot house the term go to step 7. Otherwise go to step 8.

STEP 7: Try to house the literals of the product terms in the lattice path with the permutation of the literal arrangements. The permutation will go on until the product term is housed properly or all the permutation order have been tried already. If it is not possible to house the term any way go to STEP 10.

STEP 8: If the attempt was successful in STEP 6 or STEP 7, check if there is any x x' literals in the path. Eliminate any path containing x x' literals. Also check if the path was already completed by any other path. After these checking go for the next product term for matching at STEP5.

STEP 9: If all the Product terms checking is done, then try to fix any dangling paths of the Boolean implementation. Check the following things:

- Check to see if the path is self-cancelling
- Check to see if any path is a superset of any other given term
- Check to see if any path can be fixed by setting a variable to 0
- Check if there is an unassigned variables, zero it and also mark all paths it participates. Eliminate further paths due to this zeroing.

STEP 10: In this step we have to check if the solution was found. If we get the desired solution then the process will end. And if there is no solution then we will try with next examination order from STEP 3 if any order still remaining to check. If all the orders were already checked the process will also end. Then the term has no solution with our given lattice.

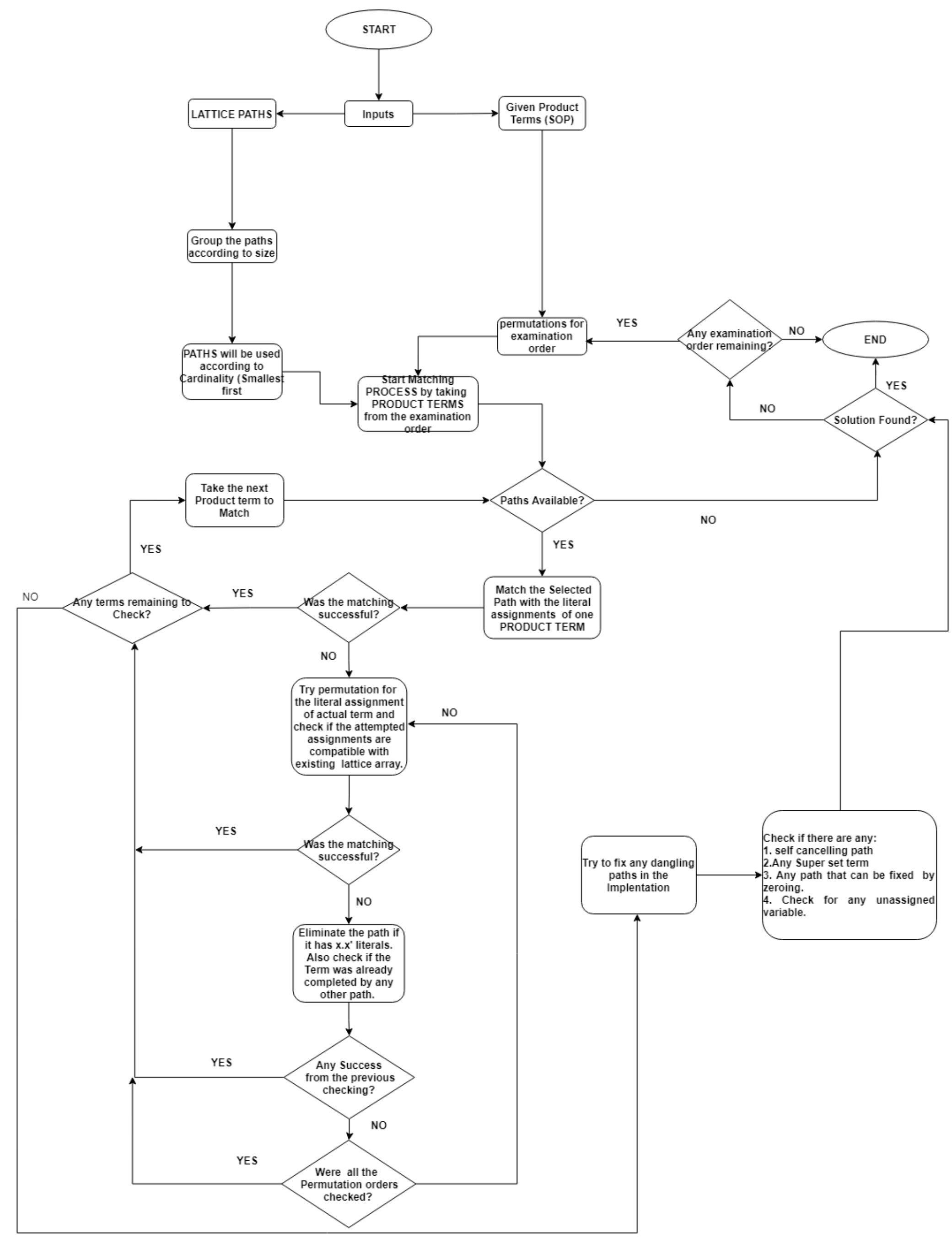

Fig 18: Flow chart of the process of the Mapping tool

### 3.3 Examples From the Mapping tool

Example 1:

Let's take an Example which we need to map in a 3 by 3 lattice.

Function: : D' B' + D' F E C' + A' F

The format of the function to give as input to the MAPPER is as follow:

3
2 997 999
4 997 5 4 998
2 1000 5

There are 3 product terms in the function.

TERM 1 : D' B'
TERM 2 : D' F E C'
TERM 3 : A' F

The MAPPER will take every term one by one and then try to place the literals of the terms in the lattice. It will check different combinations of the literals in lattice network until it gets the proper lattice assignment. If the solution is not possible it will give us "NO Solution" as answer. It means the function cannot be mapped with the given lattice. Following is the output of our MAPPER after giving the function as input.

Lattice solution:

| | | |
|---|---|---|
| E | B' | A' |
| C' | D' | F |
| 0 | B' | 1 |

Fig 19: Lattice implementation of example 1

This solution has the following point of interests:

- Term 1 covered escape path by picking multi options.

- Zero on lattice variable  6.
- Term examination order  2  1  3 .

First we can check the examination order of the solution. The mapping tool will check different orders with the terms to house the product terms in the lattice.  Here,  term examination order is 2  1  3 for the successful implementation of the function.

At first term 2 has been placed in the lattice. And it generates the term successfully.

TERM 2:  D'  F  E  C'

| E | B' | A' |
|---|---|---|
| C' | D' | F |
| 0 | B' | 1 |

Fig 20:  Lattice implementation of example 1: TERM 2

Here Top to Bottom connections are placed by     E -> C' -> D' -> F -> 1.   '1'   is placed to connect the term with the Bottom plate. It will not make any change of the term.  So this is one of the way we can place  TERM 2 in the lattice network.

TERM 1   : D'  B'

| E | B' | A' |
|---|---|---|
| C' | D' | F |
| 0 | B' | 1 |

Fig 21:  Lattice implementation of example 1: TERM 1

Here Top to Bottom connections are placed by B'->D'->B'

As we discussed in earlier chapters , if we have multiple same literals in a PATH it will count as single literal in term formation. So this path is basically B'D'.

This also could be placed as B' -> D' -> 1.

But, there are now three more paths

E C' D' B'

B' D' F 1

B' D' F A'

All these terms contains B'D'. So these are superset of B'D'. These paths will not be generated.

Here we can observe one of our point of interest:

Term 1 covered escape path by picking multi options.

So there are escape paths (paths which are redundant) which were covered by TERM 1 in multiple ways.

TERM 3 : A' F

| E | B' | A' |
|---|---|---|
| C' | D' | F |
| 0 | B' | 1 |

Fig 22: Lattice implementation of example 1: TERM 3

Here Top to Bottom connections are placed by A'->F->1 . '1' is placed to connect the term with the Bottom plate. It will not change the term. Also another path could have been created as A' F D' B'. But it is superset of A' F. So it was not generated.

We have already got all our desired TERMS. So, the MAPER has placed "0" at the 6th lattice position. It will not create any paths. The POI here is : Zero on lattice var 6

Example 2:
4

2 997 4

2 997 999

2 996 999

2 0 4

TERM 1 : D' E

TERM 2 : D' B'

TERM 3 : E' B'

TERM 4 : A E

LATTICE solution:

| E | B' | A |
|---|---|---|
| D' | 1 | E |
| 1 | E' | A |

Fig 23: Lattice implementation of example 2

This solution has the following point of interests:

- Term 1 saved escape path
- Term 4 covered escape path by picking multi options
- Term examination order 1 2 3 4
- The Examination order of this function is 1 2 3 4.

TERM 1 : D' E

| E | B' | A |
|---|---|---|
| D' | 1 | E |
| 1 | E' | A |

Fig 24: Lattice implementation of example 2: Term 1

In the lattice we can see two paths which consists of TERM 1.

E D'

E D' 1 E A

Second path will not be generated because it a Superset of the first one. Only the path marked in green will be generated. So we can say ,

Term 1 saved escape (redundant)path.

TERM 2 : D' B'

| E | B' | A |
|---|---|---|
| D' | 1 | E |
| 1 | E' | A |

Fig 25: Lattice implementation of example 2: Term 2

Path is created as B' -> 1-> D' -> 1

Here literal 1 is just a pass to the next step . So basically the Term is B'D'

TERM 3 : E' B'

| E | B' | A |
|---|---|---|
| D' | 1 | E |
| 1 | E' | A |

Fig 26: Lattice implementation of example 2: Term 3

B'E'  is generated  following the Path indicated in the figure.

Another path is created here:

B' -> 1 -> E -> A     =  B' E A

But we have another TERM in our function as  A E. So it is basically a superset. It will not be created for the TERM 4.

TERM 4 : A E

Term 4 is generated by  the path in following figure.

| E | B' | A |
|---|---|---|
| D' | 1 | E |
| 1 | E' | A |

Fig 27:  Lattice implementation of example 2: Term 4

Here the path is mapped  with

A -> E -> A

= A E

There  are other paths like

A E D'

B' E A

E D' E A

But all these paths are superset of TERM 4 : A E

So, TERM 4 covered escape  paths  and it has multiple options to do that.

So we can say:   Term 4 covered escape path by picking multi option

Example 3:

3 4 1000 1

4 4 1000 998 997

3 0 998 997

1 996

TERM 1 : E A' C' D'

TERM 2 : E A' B

TERM 3 : A C' D'

TERM 4 : E'

LATTICE solution**:**

| E | A | E' |
|---|---|---|
| A' | C' | E' |
| B | D' | E' |

Fig 28: Lattice implementation of example 3

This solution has the following point of interests:

- Path 1 saved by xx' at the end .
- Term examination order 1 2 3 4.

  TERM 1 : E A' C' D'

| E | A | E' |
|---|---|---|
| A' | C' | E' |
| B | D' | E' |

Fig 29: Lattice implementation of example 3 : TERM 1

TERM 2 : E A' B

| E | A | E' |
|---|---|---|
| A' | C' | E' |
| B | D' | E' |

Fig 30: Lattice implementation of example 3 : TERM 2

TERM 3 : A C' D'

| E | A | E' |
|---|---|---|
| A' | C' | E' |
| B | D' | E' |

Fig 31: Lattice implementation of example 3 : TERM 3

TERM 4 : E'

| E | A | E' |
|---|---|---|
| A' | C' | E' |
| B | D' | E' |

Fig 32: Lattice implementation of example 3 : TERM 4

Here in the lattice we can observe that E A' C' E' was not generated . If a path consists X X' formation that path cannot be generated. Here we can see, E A' C' E' consists of E E'. So this path will be canceled. So here we can say,

Path 1 saved by xx' .

Example 4:

3 995 3 1000

2 995 4

2 1000 4

3 995 997 999

TERM 1 : F' D A'

TERM 2 : F' D' B'

TERM 3 : A' E

TERM 4 : F' E

Lattice solution:

| F' | A' | F' |
|---|---|---|
| D | E | D' |
| A' | E | B' |

Fig 33: Lattice implementation of example 4

The solution has following point of interests:

- Term 3 covered escape path by picking multi options
- Term 4 was present but hiding
- Term examination order 1 2 3 4

TERM 1 : F' D A'

| F' | A' | F' |
|---|---|---|
| D | E | D' |
| A' | E | B' |

Fig 34: Lattice implementation of example 4: TERM 1

TERM 2 : F' D' B'

| F' | A' | F' |
|---|---|---|
| D | E | D' |
| A' | E | B' |

Fig 35: Lattice implementation of example 4: TERM 2

TERM 3 : A' E

| F' | A' | F' |
|---|---|---|
| D | E | D' |
| A' | E | B' |

Fig 36: Lattice implementation of example 4: TERM 3

TERM 3 is covered by the green line given in the figure. But if we observe there are two other paths which also covers A' E.

A' E D

A' E D' B'

These are basically escape path of A' E. These will not be generated as they are superset of A' E. So there are multiple options by which A'E can cover escape paths. Here we can say, Term 3 covered escape path by picking multi options.

TERM 4 : F' E

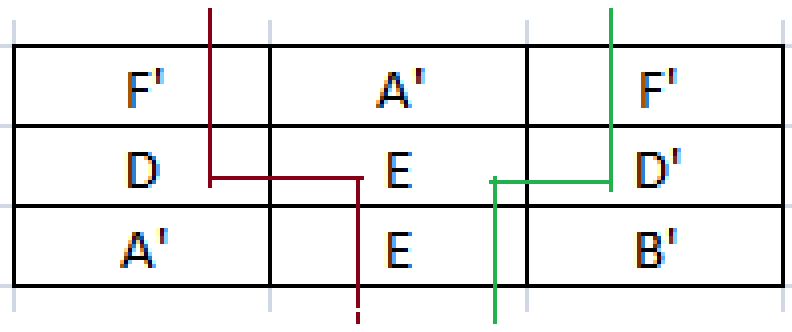

| F' | A' | F' |
|---|---|---|
| D | E | D' |
| A' | E | B' |

Fig 37: Lattice implementation of example 4: TERM 4

We have two paths mentioned here.

F' D E

F' D' E

Combining these two we can write: F' D E+ F' D' E = F'E . So these two paths actually cover F'E . TERM 4 is not present here directly but it exists in the lattice format. Here we can say, Term 4 was present but hiding.

With the above examples we have observed various ways the MAPPER can map a Boolean function. Following are all our observing points :

- Term can save escape path.
- Path can be saved by xx' format (x is literal).
- Term can covered escape path by picking multi-options.
- Zero can be used as lattice variable.
- Any Term can present in the lattice in hidden form.
- Term examination order can be different for picking the correct solution.

## CHAPTER 4

## IMPLEMENTATION OF A LATTICE NETWORK WITH SUBNETWORKS

From the analysis of function mapping in lattice networks, we can observe that some bigger networks can be implemented with a smaller network of switching lattice. Again, in few cases, it is not implementable too. With our Mapping technique, it is possible to find out if a function can be implemented with sub-functions or not.

4.1 Function implementation with sub-function:

Function with even product term: C' E' B A' + D B E + D E' B' + C' E' B' + D B A' + D A E + D A B' + C' D

SOLUTION:

| D | C' | D | D |
|---|---|---|---|
| A | C' | E' | B |
| D | D | 1 | B |
| C' | E | B' | A' |

Fig 38: Function implementation: C'E'BA' +DBE +DE'B' + C'E'B' + DBA' + DAE + DAB' + C'D

The given function can be implemented by the above lattice arrangement of 4 by 4 lattice network. Now we will try to implement this same function with two sub functions in two 3 by 3 lattice network: There can be (2^n) sub functions in this case. We will try with sub functions that carries (n/2) numberof Product terms in each. Total sub functions of (n/2) number of product terms : 70

Sub function 1:

D E' B' + D B E +D B A' +C' D

SOLUTION FOUND:

ASSG v0=E' v1=E v2=A' v3=D v4=D v5=B v6=B' v7=C' v8=D

| | | |
|---|---|---|
| E' | E | A' |
| D | D | B |
| B' | C' | D |

Fig 39: Function implementation: D E' B' + D B E +D B A' +C' D

Term 4 was placed by xx'

Term examination order 1 2 3 4

Sub function 2:

C' E' B A' + C' E' B' +D A E +D A B'

SOLUTION FOUND:

ASSG v0=D v1=C' v2=0 v3=A v4=E' v5=B v6=E v7=B' v8=A'

| | | |
|---|---|---|
| D | C' | 0 |
| A | E' | B |
| E | B' | A' |

Fig 40: Function implementation: C' E' B A' + C' E' B' +D A E +D A B'

Term 4 was present but hiding

Zero on lattice var 2

Term examination order 1 2 3 4

Adding this two sub functions, we can actually have the function which had 8 product of terms andwhich was implemented by 4 by 4 lattice.

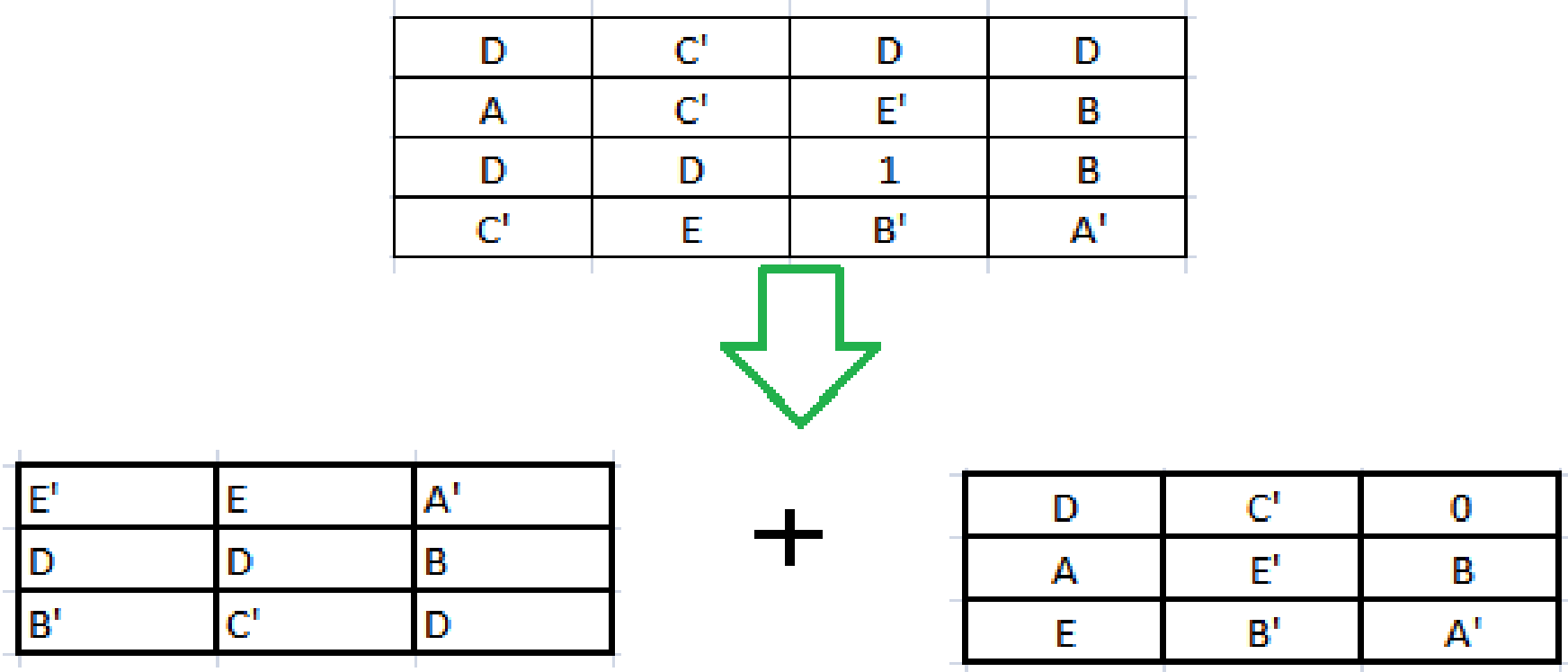

| D | C' | D | D |
|---|---|---|---|
| A | C' | E' | B |
| D | D | 1 | B |
| C' | E | B' | A' |

| E' | E | A' |
|---|---|---|
| D | D | B |
| B' | C' | D |

+

| D | C' | 0 |
|---|---|---|
| A | E' | B |
| E | B' | A' |

Fig 41: Implementation of a function with two equal size sub function in smaller network

Not only the functions which have even product terms but also those functions which have odd productterms can be split in two separate sub function to be implemented. For Example , we can take a function which has 7 product terms:

Function with odd product term: A' B' C E' + A' B' C D' + B C D' E' + B C D' A + E D' C' + E D'+BA'

It can be implemented by 4 by 4 network:

| B | E | B | A' |
|---|---|---|---|
| A' | D' | C | B' |
| 1 | 1 | D' | C |
| A' | A | C' | E' |

Fig 42: Function implementation: A'B'CE'+A'B'CD'+BCD'E'+BCD'A +ED'C'+ ED'+BA'

Now we can take the sub functions and check which two sub function can be implemented with two 3x3 lattice network to implement this function. Total sub function possible in this case: 2^7 = 128

Sub function 1:

Here the first sub function has 3 product terms:

4 1000 999 2 996

4 1000 999 2 997

4 1 2 997 996

A' B' C E' + A' B' C D' + B C D' E'

| B' | D' | 0 |
|---|---|---|
| E' | C | A' |
| B | 0 | B' |

Fig 43: Function implementation: A' B' C E' + A' B' C D' + B C D' E'

Sub function 2:

The second sub function has 4 product terms:

4 1 2 997 0

2 1 1000

2 4 997

3 4 997 998

B C D' A + E D' C' + E D' +B A'

| C | E | A' |
|---|---|---|
| A | D' | B |
| 0 | 1 | 1 |

Fig 44: Function implementation: B C D' A + E D' C' + E D' +B A'

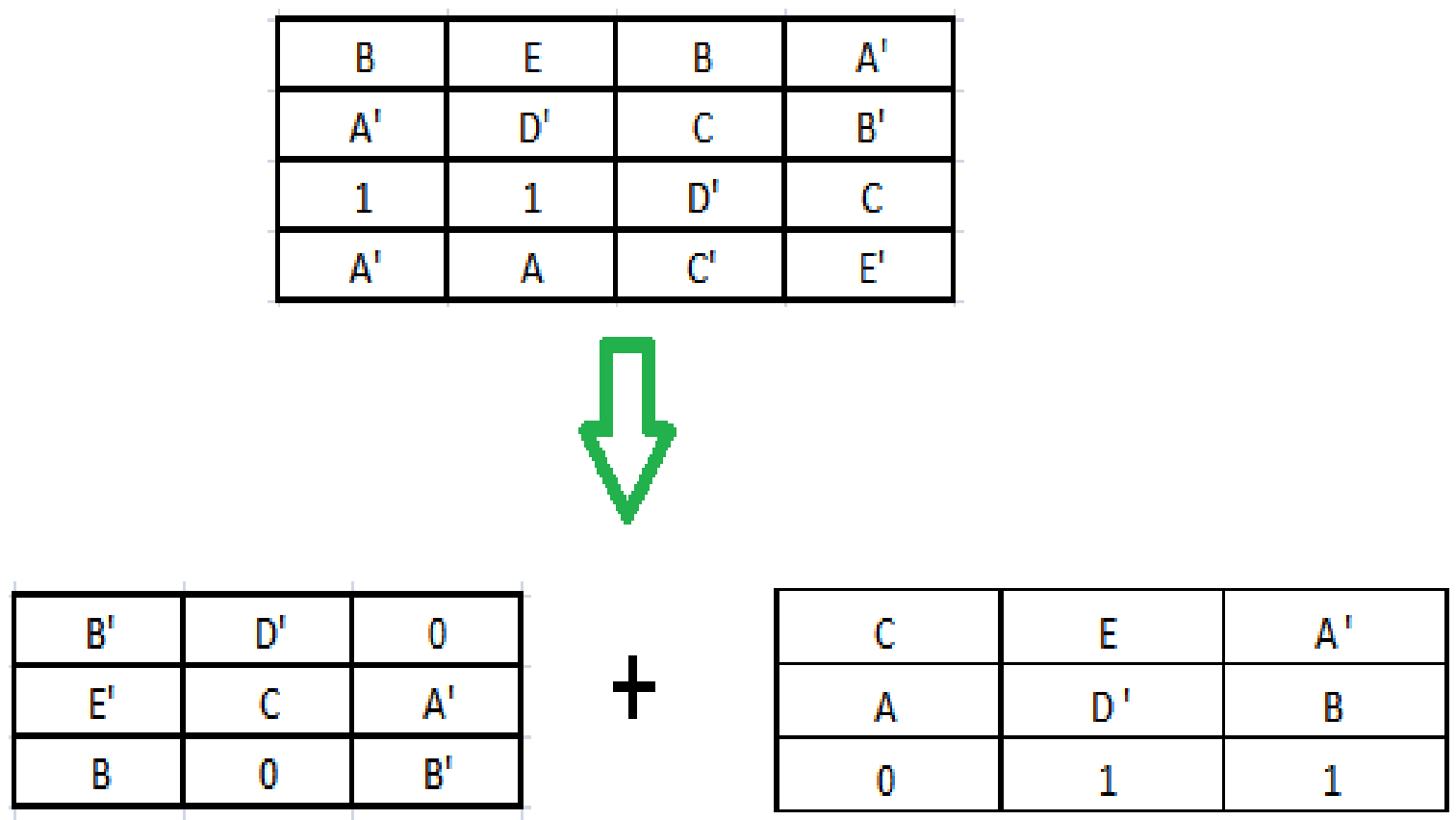

| B | E | B | A' |
|---|---|---|---|
| A' | D' | C | B' |
| 1 | 1 | D' | C |
| A' | A | C' | E' |

| B' | D' | 0 |
|---|---|---|
| E' | C | A' |
| B | 0 | B' |

+

| C | E | A' |
|---|---|---|
| A | D' | B |
| 0 | 1 | 1 |

Fig 45: Function implementation: A'B'CE'+A'B'CD'+BCD'E'+BCD'A +ED'C'+ ED'+BA' with 2 3x3 network.

Above two Example s represent the concept that , we can divide any bigger function that can beimplemented by smaller lattice network. With our Mapping technique it is possible to find out the combination for proper implementation.

**Non Implementable with two sub function (Any size ):**

All functions can't be implemented with the smaller lattice network:

Function:

EADC'B +D E'AB'C+ED'B'C +D'AB'C+A'DC'B +DE'C'B+EAB'C+ED' A'

3 4 997 1000

4 4 997 999 2

4 4 0 999 2

5 4 0 3 998 1

4 997 0 999 2

4 1000 3 998 1

4 3 996 998 1

5 3 996 0 999 2

| D | A' | D' | E |
|---|---|---|---|
| E' | D | A | 1 |
| C | C' | B' | D' |
| C' | B | C | A' |

Fig 46: Function implementation: EADC'B +D E'AB'C+ED'B'C +D'AB'C+A'DC'B +DE'C'B+EAB'C+ED' A'

There are 256 sub function possible for this function. But choosing any of the two sub functions it is notpossible to implement this function with two 3x3 lattice network.

**<u>Not implementable with equal size sub function but implementable with non equal size sub function:</u>**

There may be some cases where the function has even number of product terms but cannot be implemented by equally sized divided sub functions, rather it can be implemented by two not equalsize sub function.

Function**:**

B E' C' D' + B E' C' A' + B E' D A' + E D A' + E D B' + C E B' + B' A D' +B' A E

4 1 996 998 997

4 1 996 998 1000

4 1 996 3 1000

3 4 3 1000

3 4 3 999

3 2 4 999

3 999 0 997

3 999 0 4

| B' | C | E | B |
|---|---|---|---|
| A | E | D | E' |
| D' | B' | A' | C' |
| 0 | B' | 1 | D' |

Fig 47: Function implementation: BE'C'D' + BE'C'A' + BE'DA' + EDA' +EDB' + CEB' + B'AD' +B'AE

If we search all the sub functions which has 4 product terms to implement this function with two latticenetwork , we can't find any solution.

Rather if we try for unequal size sub function then we can have a solution:

Sub function 1:
3

4 1 996 998 997

4 1 996 998 1000

4 1 996 3 1000

B E' C' D' + B E' C' A' + B E' D A'

| D' | 0 | B |
|---|---|---|
| B | E' | A' |
| D | C' | 0 |

Fig 48: Function implementation: B E' C' D'+ B E' C' A'+ B E' D A'

Sub function 2 :
5

3 4 3 1000

3 4 3 999

3 2 4 999

3 999 0 997

3 999 0 4

E D A' + E D B' + C E B' + B' A D' +B' A E

| A | C | E |
|---|---|---|
| B' | E | D |
| D' | B' | A' |

Fig 49: Function implementation: E D A' + E D B' + C E B' + B' A D' +B' A E

So the function can be implemented by a 4x4, and it can also be implemented by two 3x3 lattices for not equal sizes subsets, but cannot be implemented by the equal size subsets.

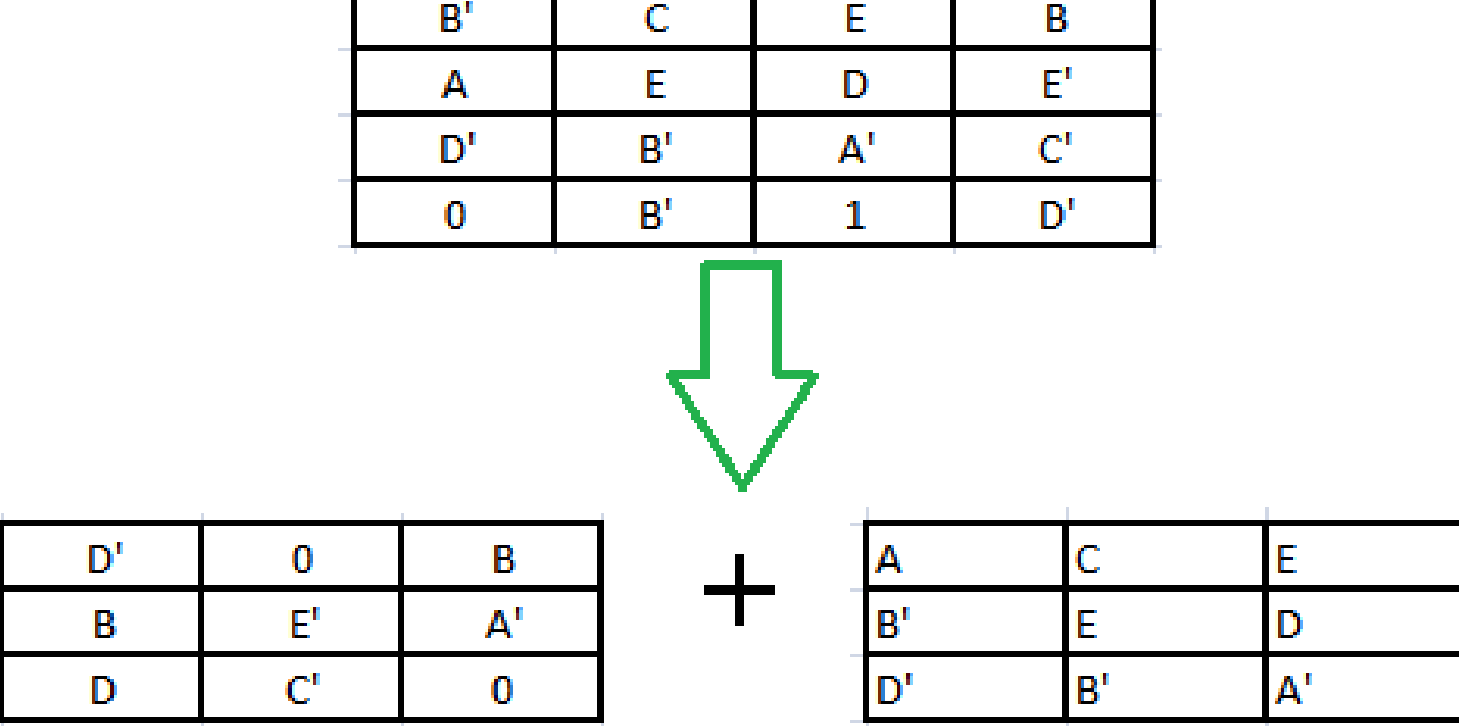

| B' | C | E | B |
|---|---|---|---|
| A | E | D | E' |
| D' | B' | A' | C' |
| 0 | B' | 1 | D' |

| D' | 0 | B |
|---|---|---|
| B | E' | A' |
| D | C' | 0 |

+

| A | C | E |
|---|---|---|
| B' | E | D |
| D' | B' | A' |

Fig 50: Function implementation BE'C'D' + BE'C'A' + BE'DA' + EDA' + EDB' + CEB' + B'AD' +B'AE with 2 3x3 lattice

# CHAPTER 5

## SYSTEMATIC PROCEDURE OF MAPPING

In this chapter we will present a systematic procedure by which we can implement any Boolean function with most efficient way by any given four terminal switching network. It will take minimum number of lattice to implement the function. For doing that first we will propose an algorithm named as 'Sub-function Algorithm'.

### 5.1 Sub-function algorithm

The process of finding the solution for any function with two lattice networks will be called Sub-function algorithm. Let's consider a function that cannot be mapped with a single lattice network (for example one 3 x3 lattice). But we may implement it with two lattice networks.

Let's,

n = Number of Product terms in a function

n/2 = half of the number of product terms

*keven*= n/2 -1 (for the function having even product term)

*kodd* = (n+1)/2 -2 (for the function having odd product term)

For 'n' number of product terms in a function:

We will make pair of sub function of every possible size to check the best possible mapping solution.

1. Start making the first pair of sub function by taking ( n/2+*keven*) number of product terms (for the function which has even product term) and (n+1)/2+*kodd* number of product terms (for the function which has odd product term).

<u>For example</u>**:**

**1.** If in a function there are 8 product terms, we will first take a sub function of size consisting $n/2+keven = 4+3 = 7$ product terms. (here, $n/2 = 4$, $keven = n/2 - 1 = 3$) .

If in a function there are 9 product terms, we will first take a sub function of size consisting $(n+1)/2 + kodd = 5 + 3 = 8$ product terms. (here, $n+1/2 = 5$, $kodd = (n+1)/2 - 2 = 5 - 2 = 3$).

**2.** Then create the other sub function of the pair by taking **:**

$n - (n/2+keven)$ number of the product terms that were **unused** in the first sub function [ for even number of product terms]

$n - [(n+1)/2 + kodd]$ number of the product terms that were **unused** in the first sub function [ for odd number of product terms]

From the Example of previous step:

we will take $n - (n/2+keven) = 8-7 = 1$ remaining terms to make the another sub function for even sample and $n - [(n+1)/2 + kodd] = 9 - 8 = 1$ remaining terms to make the another sub function for the odd sample. We will create all possible sub function of these sizes in pairs.

**3.** After creating the all the pairs starting from $n/2+keven$ (for odd: $(n+1)/2 + kodd$ ) and its correspondent sub function $n - (n/2+keven)$ (for odd: $n - [(n+1)/2 + kodd)]$) , go for the pair of subset of size consisting of :

" $n/2+keven - x$ " ( for odd: $(n+1)/2 + kodd - x$) and

" $n-(n/2+keven)+x$ " ( for odd: $n - (n+1)/2 - kodd + x$)

where $x = 1, 2, 3.... .n/2$ [ for odd: $(n+1)/2$ ]

The function which has 8 product term in our previous example, first we created all the pair of sub-function with 7 product term and sub-function with 1 product term.

Now in this step we will create the next pair :

$n/2 + \mathit{keven} - x = 4 + 3 - 1 = 6$

and $n - ( n/2+\mathit{keven} ) + x = 8 - 7 + 1 = 2$

[where x=1]

Produce all pairs of 6 product terms and 2 product terms.

In the same way for the function with 9 product terms in our previous example, first, we created the sub-function of product terms: 8 product terms and 1 product term.

Now in this step we will create the next pair of sub function :

$(n+1)/2 + \mathit{kodd} - x = 5+ 3 -1 = 7$

$n - [(n+1)/2+\mathit{kodd}] + x = 9 - 8 + 1 = 2$

[where x=1]

Produce all pairs of 7 product term and 2 product term.

Thus we can create pairs for all values of x

*where x= 1, 2, 3.....n/2 [For even ]*

*where x= 1, 2, 3..... (n+1)/2 [For odd]*

The sub function creation can be stopped at any predetermined value of x . This whole process we will call as "SUB FUNCTION algorithm" .

The order of creating pairs for sub-function by this algorithm will help to find out the best possible match of **the maximum number of product terms** that any lattice network can map for that function. When we will try to find any pair that can be implemented by 2 lattice networks, we will take the first pair from the SUB FUNCTION lists that can be mapped. We don't need to check the others after we get our first solution. Whenever we get our solution, we will be sure that one of the sub-functions (which has larger product terms) has been implemented by using the **maximum possible product terms** mapped in a lattice network. The second lattice or the smaller one **may** map more product terms than it has.

For example:

if any function has 8 product terms, we can create pairs of sub-function as follow:

7 product term and 1 product term

6 product term and 2 product term

5 product term and 3 product term

4 product term and 4 product term

Suppose, we got our first mapping solution with :

6 product terms and 2 product terms pair.

So from this, we can say, one lattice network can implement a maximum of 6 product terms from the given function. Because we checked with 7 product terms and there was no solution. So the largest number of product terms that our given lattice network can map is 6 here. We can take the lattice network with the 6 product term as one of the solutions. The smallest sub-function which has 2 product terms may implement more than 2.

### 5.2 Flowchart of the Process

Definitions**:**

**Ln**: Product terms of given function

**Lni :** Individual input product terms

**Lny:** Largest part after splitting

**LB**: Longest Product term of basic lattice network

**N**: Number of product term in input function

In the following page the flow chart of the procedure is shown.

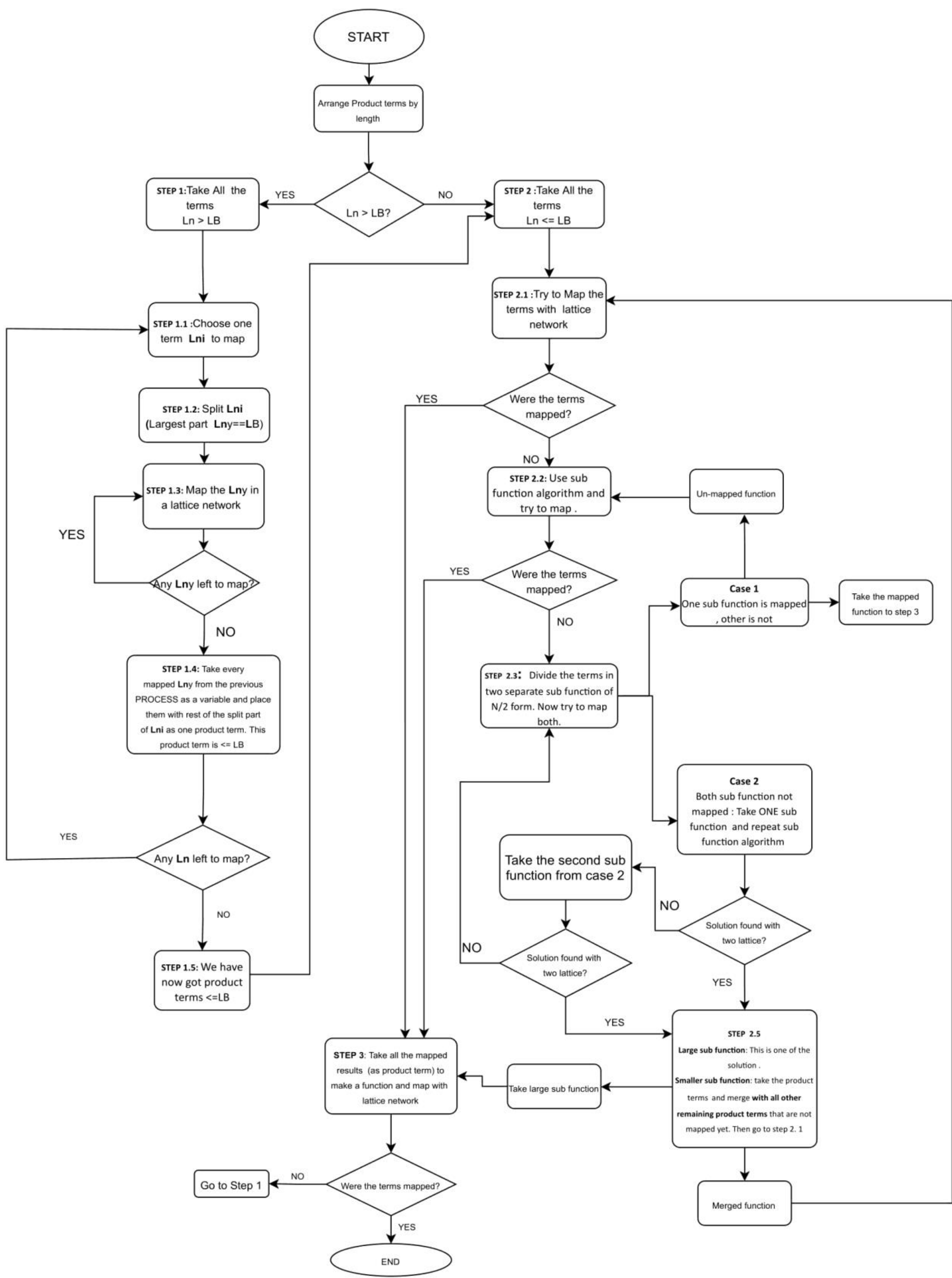

Fig 51 : Flowchart of the systematic procedure

## 5.3 Example of Mapping of any Boolean function with 3x3 lattice

Example 1:

Let's take an Example q = abcdefg + ab'e + a'cdf'

According with our Flowchart:

1. First we will arrange all the terms according to their length.

2. Check if there are any terms (**Ln**) longer than the largest term (**LB**)of 3 by 3 basic lattice paths (longest path in a 3 by 3 basic lattice is 5). Also need to check how many terms are there like this.

3. Then we can take these terms (**Ln** > **LB**) one by one and split each into several terms. For splitting a product term (**Lni**) into several terms, the highest length (**Lny** )of a new term should not exceed the length of **LB**.

For our Example , we will split the first term and make a new sub function.

Ln1 = a b c d e f g = a b c d e + f g

split terms are:

x1= a b c d e

x2 = f g

As the output from the lattice will be in the SUM format we will need to Map each split term with a 3x3 lattice

| | | |
|---|---|---|
| a | 0 | 0 |
| b | c | d |
| 0 | 0 | e |

Fig 52: x1= a b c d e

Now if we try combine the both x1 & x2 as a single product term, we will get

Product term X= f g x1

This is basically the implementation of the product term X= a b c d e f g [x1= a b c d e ]

So , now the Product term X is < = Lb [X= f g x1]

4. As there is no more term that is longer than **LB**, we will merge all the terms

f g x1

a b' e

a' c d f'

But these three terms cannot be mapped with single 3 by 3 lattice structure.

Now we will try to Map with 2 Lattice using SUB FUNCTION algorithm.

| a | 0 | f |
|---|---|---|
| b' | 0 | g |
| e | 0 | X1 |

Fig 53: Implementation of a b' e + f g x1

| a' | 0 | 0 |
|---|---|---|
| c | d | 0 |
| 0 | f' | 0 |

Fig 54: Implementation of a' c d f'

It can be implemented with 2 3x3 lattice.

So finally , we got the solution :
result x1= a b c d e

result = a b' e + f g x1 = a b' e + f g a b c d e [x1= a b c d e ]

result : a' c d f

So all the product terms of q are mapped with 3 lattice:

q = abcdefg + ab'e + a'cdf'

Example 2:

Function**:**
E D C' B + D A B' C + E D' B' C + D' A B' C + A' D C' B + D E' C' B + E A B' C + E D' A'

8
3 4 997 1000
4 4 997 999 2
4 4 0 999 2
4 4 3 998 1
4 1000 3 998 1
4 3 996 998 1
5 3 996 0 999 2
4 997 0 999 2

We will show this Example with every step mentioned in the flow chart :

STEP 1 **:** Try to Map the Function :

result with our Example : Not possible to MAP.

STEP 2 : Use sub function algorithm and try to map them.
result with our Example : Not possible to MAP with two lattice using sub function algorithm

STEP 3: If any solution is found then our function is implemented with two lattice network.
If implementation is not possible then divide the function in two sub function of N/2 form.

result with our Example : TWO sub function of N/2 form:

4
3 4 997 1000
4 4 997 999 2
4 4 0 999 2
4 4 3 998 1

4
4 1000 3 998 1
4 3 996 998 1
5 3 996 0 999 2
4 997 0 999 2

STEP 4: Now try to implement both the function of N/2 form .

Result with our Example : No solution for any of the function. So we will approach to (case2)

Case2: If there is a case where both sub function was not mapped then we take one by one sub function and again goes for the sub function algorithm we mentioned at step 2.

First take the following sub function:
4
3 4 997 1000
4 4 997 999 2
4 4 0 999 2
4 4 3 998 1

By using sub function algorithm we can check if we get solution as a pair of sub functions.

3
3 4 997 1000
4 4 997 999 2
4 4 0 999 2

SOLUTION FOUND:

ASSG v0=E v1=0 v2=E v3=D' v4=B' v5=A v6=A' v7=C v8=0

We have got solution for the largest sub function between two sub function.

The last product term here is a separate sub function. We can easily map it.

4 4 3 998 1

SOLUTION FOUND:

ASSG v0=E v1=0 v2=v3=D v4=C' v5=0 v6=0 v7=B v8=0

STEP 5.
Case 1: From Step 4 (Case2) If any combination pair was found which were mapped, then take the function which is LARGEST in that pair. That is one of our lattice solution network.

Now take the product terms of the smallest function of the pair and merge with the other remaining sub function from the step 3.

We will combine the smaller sub function from STEP 4 with the remaining sub function from the step 3.

result after the merger**:**

5
**4 4 3 998 1** (smallest Sub function from STEP 4)
4 1000 3 998 1
4 3 996 998 1
5 3 996 0 999 2
4 997 0 999 2

STEP 6 : Start the process again from STEP 1 until we map all the product terms.

5
4 4 3 998 1
4 1000 3 998 1
4 3 996 998 1
5 3 996 0 999 2
4 997 0 999 2

Implement of the function is not possible with a single 3x3 lattice. So we need to check with sub function algorithm.

Largest sub function of the pair which maps:
3

4 4 3 998 1

4 1000 3 998 1

4 3 996 998 1

SOLUTION FOUND:

ASSG v0=E v1=0 v2=D v3=D v4=C' v5=E' v6=0 v7=B v8=0

The smallest sub function:
2

5 3 996 0 999 2

4 997 0 999 2

We don't have any other sub function left . So we have to try to map only this now.

5 3 996 0 999 2

4 997 0 999 2

SOLUTION FOUND:

ASSG v0=D v1=D' v2=0 v3=E' v4=A v5=B' v6=0 v7=0 v8=C

So finally We got our solution in 3x3 network using lowest possible lattice:
3

3 4 997 1000

4 4 997 999 2

4 4 0 999 2

SOLUTION FOUND:

ASSG v0=E v1=0 v2=E v3=D' v4=B' v5=A v6=A' v7=C v8=0

4 4 3 998 1

4 1000 3 998 1

4 3 996 998 1

SOLUTION FOUND:

ASSG v0=E v1=0 v2=D v3=D v4=C' v5=E' v6=0 v7=B v8=0

5 3 996 0 999 2

4 997 0 999 2

SOLUTION FOUND:

ASSG   v0=D v1=D' v2=0 v3=E' v4=A v5=B' v6=0 v7=0 v8=C

# CHAPTER 6

# CONCLUSION AND FUTURE WORK

## 6.1 CONCLUSION

In this work, we presented design techniques to synthesize Boolean Function with four terminal switching networks. Three steps were mainly shown in this work. First, we represented the design technique to generate a library of functions. Generation of functions is helpful to realize the circuit implementation with lattice network. These functions can be used for test purposes. We also discussed how to use that design to use as a lattice network solver.

In the second part, we have discussed the mapping tool for synthesizing functions in a lattice network using the paths of that network. This synthesizing tool is very important for the switching lattice network implementation as many functions can be very hard to implement in the four-terminal lattice network due to their connection complexity. If the lattice size increased the paths of the lattice also increase.

Finally, we have shown a systematic way to implement any function with any given lattice network. In this case, we have considered using the same kind of lattice network for systematic implementation.

## 6.2 FUTURE WORK

In our work, we have developed the mapping technique using the lattice paths. In some other work, this kind of mapping technique was done using a minimization tool and SAT solver [10] [13]. We can compare both techniques to achieve more efficiency in the tool implementation. Our design can handle up to 6x6 formation of lattice till now, which is already a very large lattice network with a huge number of paths. In future efforts, it is possible to make the tool efficient to synthesize functions with larger lattices. The implementation of the systematic procedure we discussed in chapter 5 is also a good scope for future works.

## REFERENCES

[1] Dubash, M. Moore's Law is dead, says Gordon Moore. Techworld. com, 13 (2005).

[2] Altun, Mustafa, and Marc D. Riedel. "Logic synthesis for switching lattices." Computers, IEEE Transactions on 61.11 (2012): 1588-1600.

[3] C.E. Shannon, "A Symbolic Analysis of Relay and Switching Circuits," Trans. Am. Inst. of Electrical Eng., vol. 57, no. 12, pp. 713-723, 1938.

[4] S.B. Akers, "A Rectangular Logic Array," IEEE Trans. Computers, vol. C-21, no. 8, pp. 848-857, Aug. 1972.

[5] M. Chrzanowska-Jeske, Y. Xu, and M. Perkowski, "Logic Synthesis for a Regular Layout," VLSI Design, vol. 10, no. 1, pp. 35-55, 1999.

[6] M. Chrzanowska-Jeske and A. Mishchenko, "Synthesis for Regularity Using Decision Diagrams," Proc. IEEE Int'l Symp. Circuits and Systems (ISCAS), pp. 4721-4724, 2005.

[7] M.M. Ziegler and M.R. Stan, "CMOS/Nano Co-Design for Crossbar-Based Molecular Electronic Systems," IEEE Trans. Nanotechnology, vol. 2, no. 4, pp. 217-230, Dec. 2003.

[8] Y. Zomaya, "Molecular and Nanoscale Computing and Technology," Handbook of Nature-Inspired and Innovative Computing, ch. 14, pp. 478-520, Springer, 2006.

[9] S. Safaltin et al., "Realization of Four-Terminal Switching Lattices: Technology Development and Circuit Modeling," 2019 Design, Automation & Test in Europe Conference & Exhibition (DATE), 2019, pp. 504-509, doi: 10.23919/DATE.2019.8715123.

[10] M. Ceylan Morgül, Mustafa Altun, Optimal and heuristic algorithms to synthesize lattices of four-terminal switches, Integration, Volume 64, 2019, Pages 60-70, ISSN 0167-9260, https://doi.org/10.1016/j.vlsi.2018.08.002.

[11] Anna Bernasconi, Valentina Ciriani, Luca Frontini, Gabriella Trucco, Composition of

switching lattices for regular and for decomposed functions, Microprocessors and Microsystems, Volume 60, 2018, Pages 207-218,ISSN 0141-9331, https://doi.org/10.1016/j.micpro.2018.05.004.

[12] A. Bernasconi, A. Boffa, F. Luccio and L. Pagli, "Two Combinatorial Problems on the Layout of Switching Lattices," 2018 IFIP/IEEE International Conference on Very Large Scale Integration (VLSI-SoC), 2018, pp. 137-142, doi: 10.1109/VLSI-SoC.2018.8644855.

[13] L. Aksoy and M. Altun, "A Satisfiability-Based Approximate Algorithm for Logic Synthesis Using Switching Lattices," 2019 Design, Automation & Test in Europe Conference & Exhibition (DATE), 2019, pp. 1637-1642, doi: 10.23919/DATE.2019.8714809.

[14] N. Akkan et al., "Technology Development and Modeling of Switching Lattices Using Square and H Shaped Four-Terminal Switches," in IEEE Transactions on Emerging Topics in Computing,

[15] Bernasconi A., Boffa A., Luccio F., Pagli L. (2019) The Connection Layout in a Lattice of Four-Terminal Switches. In: Bombieri N., Pravadelli G., Fujita M., Austin T., Reis R. (eds) VLSI-SoC: Design and Engineering of Electronics Systems Based on New Computing Paradigms. VLSI-SoC 2018. IFIP Advances in Information and Communication Technology, vol 561. Springer, Cham. https://doi.org/10.1007/978-3-030-23425-6_3

[16] A. Bernasconi, V. Ciriani and L. Frontini, "Testability of Switching Lattices in the Stuck at Fault Model," 2018 IFIP/IEEE International Conference on Very Large Scale Integration (VLSI-SoC), 2018, pp. 213-218, doi: 10.1109/VLSI-SoC.2018.8644806.

[17] L. Aksoy and M. Altun, "Novel Methods for Efficient Realization of Logic Functions Using Switching Lattices," in IEEE Transactions on Computers, vol. 69, no. 3, pp. 427-440, 1 March 2020, doi: 10.1109/TC.2019.2950663.

# VITA

Graduate School
Southern Illinois University

Rajesh Kumar Datta

rajeshcuet10@gmail.com

Chittagong University of Engineering and Technology
Bachelor of Science, Electrical and Electronic Engineering, December 2014

Thesis Paper Title:
Implementing Boolean Function With Switching Lattice Network.

Major Professor: Dr. Dimitrios Kagaris